\documentclass[twocolumn]{aastex631}

\usepackage{natbib}
\usepackage{amsmath}
\usepackage{graphicx}
\usepackage{caption}
\usepackage{cleveref}
\shorttitle{The Thermal Emission Spectrum of the Nearby Rocky Exoplanet LTT 1445A b}
\shortauthors{Wachiraphan et al.}
\graphicspath{{./}{figures/}}
\begin{document}

\title{The Thermal Emission Spectrum of the Nearby Rocky Exoplanet LTT 1445A b from JWST MIRI/LRS}

\correspondingauthor{Patcharapol Wachiraphan}
\email{patcharapol.wachiraphan@colorado.edu}

\author[0000-0001-6484-7559]{Patcharapol Wachiraphan}
\affiliation{Department of Astrophysical and Planetary Sciences, University of Colorado Boulder, Boulder, CO 80309, USA}
\affiliation{Center for Astrophysics and Space Astronomy, University of Colorado Boulder, 593 UCB, Boulder, CO 80309, USA}

\author[0000-0002-3321-4924]{Zachory K. Berta-Thompson}
\affiliation{Department of Astrophysical and Planetary Sciences, University of Colorado Boulder, Boulder, CO 80309, USA}
\affiliation{Center for Astrophysics and Space Astronomy, University of Colorado Boulder, 593 UCB, Boulder, CO 80309, USA}

\author[0000-0001-8274-6639]{Hannah Diamond-Lowe}
\affiliation{Department of Space Research and Space Technology, Technical University of Denmark, Elektrovej 328, 2800 Kgs.\,Lyngby, Denmark}
\affiliation{Space Telescope Science Institute, Baltimore, MD 21218, USA}

\author[0000-0001-6031-9513]{Jennifer G. Winters}
\affiliation{Bridgewater State University, 131 Summer St., Bridgewater, MA 02325, USA}
\affiliation{Center for Astrophysics $\vert$ Harvard \& Smithsonian, 60 Garden Street, Cambridge, MA 02138, USA}

\author[0000-0001-8504-5862]{Catriona Murray}
\affiliation{Department of Astrophysical and Planetary Sciences, University of Colorado Boulder, Boulder, CO 80309, USA}
\affiliation{Center for Astrophysics and Space Astronomy, University of Colorado Boulder, 593 UCB, Boulder, CO 80309, USA}

\author[0000-0002-0659-1783]{Michael Zhang}
\affiliation{Department of Astronomy \& Astrophysics, University of Chicago, Chicago, IL 60637, USA}

\author[0000-0002-6215-5425]{Qiao Xue}
\affiliation{Department of Astronomy \& Astrophysics, University of Chicago, Chicago, IL 60637, USA}

\author[0000-0002-4404-0456]{Caroline V. Morley}
\affiliation{Department of Astronomy, University of Texas at Austin, 2515 Speedway, Austin, TX 78712, USA}

\author[0000-0003-0216-559X]{Marialis Rosario-Franco}
\affiliation{Department of Astrophysical and Planetary Sciences, University of Colorado Boulder, Boulder, CO 80309, USA}

\author[0000-0002-7119-2543]{Girish M. Duvvuri}
\affiliation{Department of  Physics and Astronomy, Vanderbilt University, Nashville, TN 37235, USA}



\begin{abstract}
The nearby transiting rocky exoplanet LTT 1445A b presents an ideal target for studying atmospheric retention in terrestrial planets orbiting M dwarfs. It is cooler than many rocky exoplanets yet tested for atmospheres, receiving a bolometric instellation similar to Mercury's. Previous transmission spectroscopy ruled out a light H/He-dominated atmosphere but could not distinguish between a bare-rock, a high-MMW, or a cloudy atmosphere. We present new secondary eclipse observations using JWST's MIRI/LRS, covering the 5-12 $\mu$m range. From these observations, we detect a broadband secondary eclipse depth of 41 $\pm$ 9 ppm and measure a mid-eclipse timing consistent with a circular orbit (at 1.7$\sigma$). From its emission spectrum, the planet’s dayside brightness temperature is constrained to 525 $\pm$ 15\,K, yielding a temperature ratio relative to the maximum average dayside temperature from instant thermal reradiation by a rocky surface $R$ = $T_{\rm day,obs}/T_{\rm max}$ = 0.952 $\pm$ 0.057, consistent with emission from a dark rocky surface. From an energy balance perspective, such a warm dayside temperature disfavors thick atmospheres, excluding $\sim$100 bar atmospheres with Bond albedo $>$ 0.08 at the 3$\sigma$ level. Furthermore, forward modeling of atmospheric emission spectra disfavor simple 100\% CO$_2$ atmospheres with surface pressures of 1, 10, and 100 bar at 4.2$\sigma$, 6.6$\sigma$, and 6.8$\sigma$ confidence, respectively. These results suggest that LTT 1445A b lacks a very thick CO$_2$ atmosphere, possibly due to atmospheric erosion driven by stellar activity. However, the presence of a moderately thin atmosphere (similar to those on Mars, Titan, or Earth) remains uncertain.
\end{abstract}



\section{Introduction} \label{sec:intro} % 1-2 pages

The atmospheres of terrestrial planets within our solar system, including Mercury \citep{benz_origin_2007, asphaug_mercury_2014}, Venus \citep{bullock_atmosphere_2013}, and Mars \citep{owen_composition_1977, kieffer_thermal_1977}, have been extensively studied over the years. However, our understanding of the origin and evolution of atmospheres under various stellar and planetary circumstances remains limited due to the lack of a large population sample. A significant knowledge gap lies in delineating the boundary between minimal-atmosphere hot-dayside rocky planets, such as Mercury (with surface pressures on the order of 10$^{-14}$ bar), and those with thick CO$_2$ atmospheres resembling Venus (with surface pressures of 92 bar). The discovery of exoplanets orbiting main-sequence stars has opened new avenues for studying planetary populations and their atmospheres \citep{mayor_jupiter-mass_1995, charbonneau_detection_2002}, particularly those that resemble terrestrial planets in terms of radius, mass, and bulk density \citep{batalha_kepler_2011, berta-thompson_rocky_2015}. Nevertheless, investigating the atmospheres of Earth-Sun analogs remains a formidable challenge.

The opportunity to detect and study the atmospheres of terrestrial exoplanets by observing those transiting M dwarfs offers distinct advantages over other types of stars \citep{national_academies_of_sciences_exoplanet_2018}. Firstly, planets orbiting M dwarfs tend to have higher planet-to-star contrast due to the lower stellar luminosity, resulting in a higher signal-to-noise ratio (SNR) for many observables. Secondly, M dwarfs constitute approximately 75\% of all stars in our galaxy \citep{henry_solar_2006}, hence increasing the expected number of nearby transiting rocky exoplanets \citep{berta-thompson_rocky_2015, gillon_seven_2017}, which opens up more opportunities for atmospheric observations. However, the existence of atmospheres on M dwarf terrestrial exoplanets remains uncertain due to the typically high stellar activity of M dwarfs, which often lasts long after they enter the main sequence \citep[e.g.,][]{lammer_coronal_2007,france_high-energy_2020,king_euv_2020, medina_galactic_2022}. Multiple comprehensive reviews have discussed the potential for atmospheres and habitability of rocky exoplanets around M dwarfs, including those by \citet{tarter_reappraisal_2007}, \citet{scalo_m_2007}, and the more recent review by \citet{shields_habitability_2016}.

Understanding the bulk density of exoplanets, derived from extensive studies of their masses and radii, has provided valuable insights into atmospheric retention. For example, \citet{owen_evaporation_2017} demonstrated that the observed bimodal distribution of small planet sizes (approximately 1-5 $R_{\oplus}$) from the Kepler mission could result from photoevaporation of hydrogen envelopes, primarily driven by X-ray and ultraviolet (XUV) radiation from the host star. This effect may create the observed gap in the distribution at 1.8 $R_{\oplus}$. Additionally, the changing bimodal shape at different planetary ages \citep{berger_gaiakepler_2020,cloutier_evolution_2020,vanwyngarden_modeling_2024} suggests the persistence of photoevaporation effects over time. \citet{luger_habitable_2015} used the energy-limited approximation to estimate the mass-loss rate of exoplanets in the habitable zone around M dwarfs. They found that even with relatively inefficient processes (with only 15\% of the XUV flux driving atmospheric loss), H/He atmospheres of habitable exoplanets are likely to be mostly stripped away early in their histories. On longer timescales (0.5 - 1 Gyr), planetary atmospheres can still be lost due to core-powered mass loss \citep{ginzburg_core-powered_2018, gupta_signatures_2020}. However, several studies suggest that small-radius, close-in planets can acquire secondary atmospheres, as seen on Venus, Earth, and Mars, through various mechanisms, including volatile outgassing from the planet's interior \citep{tian_thermal_2009} and comet or asteroid bombardment \citep{dorn_secondary_2018}. However, the observational confirmation of such secondary atmospheres on any rocky M dwarf planets remains tenuous.

Over the years, several methods have been developed and deployed to characterize the atmospheres of transiting exoplanets. In the following, we will discuss potential approaches to investigate the atmospheres of terrestrial exoplanets orbiting M dwarfs. Transmission spectroscopy, which measures variations in transit depth across different wavelength bands \citep{brown_transmission_2001}, has long been used to study atmospheric composition and properties \citep[e.g.,][]{charbonneau_detection_2002,tinetti_water_2007,deming_infrared_2013}. For rocky planets, transmission spectroscopy has limitations in distinguishing between the muted absorption features in cloudy/hazy atmospheres and no-atmosphere bare-rock scenarios of rocky exoplanets \citep[e.g.,][]{kreidberg_clouds_2014,de_wit_combined_2016,diamond-lowe_ground-based_2023}. Phase curve observations \citep{knutson_map_2007} provide comprehensive information over the complete orbit of an exoplanet, including atmospheric heat redistribution \citep{kreidberg_absence_2019,hammond_reliable_2024}. However, they require extensive telescope time to detect variations in out-of-transit light curves since the full planetary orbit must be observed. Emission spectroscopy at secondary eclipse \citep{charbonneau_detection_2005} has emerged as a potentially promising approach to confirm or refute the presence of atmospheres on rocky exoplanets \citep{zhang_gj_2024}.

For high SNR targets such as hot Jupiters, thermal emission spectra can identify molecular abundances and thermal profiles \citep[e.g.,][]{cartier_near-infrared_2017, sheppard_evidence_2017}. However, with cooler and terrestrial-size exoplanets, thermal emission spectra might not be suitable for identifying the individual features or thermal profile due to the low SNR of the observation. For tidally-locked terrestrial planet atmospheres, the presence of an atmosphere regulates the dayside temperature \citep{joshi_simulations_1997} which can be measured through thermal emission detected at secondary eclipse \citep{morley_observing_2017,koll_identifying_2019,kreidberg_absence_2019,crossfield_gj_2022}. Without directly detecting atmospheric absorption or emission features, the presence of an atmosphere might be inferred through the dayside brightness temperature ($T_{\rm day}$), which, following \citet{burrows_spectra_2014}, can be written as
\begin{eqnarray}
    T_{\rm day} &=& T_{\ast,\text{eff}} \sqrt{\frac{R_{\ast}}{a}} (1-\alpha_{\rm B})^{1/4}f^{1/4} \label{eqn:dayside_temp_with_albedo} 
    %&=& (4f)^{1/4}T_{\text{eq}} \label{eqn:dayside_temp} 
\end{eqnarray} 
where $T_{\ast,\text{eff}}$ and $R_{\ast}$ are the stellar effective temperature and radius, respectively. $a$ is the distance of the planet from its host star. $\alpha_{\rm B}$ is planet's Bond albedo, the wavelength-integrated fraction of the total incident stellar radiation reflected by the planet. $f$ is the heat redistribution factor, which strongly correlates with the ratio between the area receiving stellar flux and the effective surface area of reradiating planet flux. For perfect uniform heat redistribution from rapid rotation or an atmosphere circulating heat, the entire planet reradiates with a surface area of $4\pi R_p^2$, while the cross-sectional area of flux received is always $\pi R_p^2$. Therefore, $f = \frac{\pi R_p^2}{4\pi R_p^2} = 1/4$. Likewise, if the planet redistributes heat uniformly over only its dayside hemisphere, the surface area would be $2\pi R_p^2$, translating to an $f$-factor of 1/2. Due to reradiation geometry weighted more toward the substellar point more than limbs, the instant reradiation has an effective surface area of $\frac{3}{2}\pi R_p^2$, hence $f = 2/3$. In this paper, we will use the nomenclature of equilibrium temperature ($T_{\rm eq}$) strictly for a temperature of planet with uniform heat redistribution ($f$-factor = 1/4) and zero Bond albedo.

\citet{koll_scaling_2022} used dayside brightness temperature as a proxy to constrain the surface pressure under simplified physical assumptions. This approach bridges the gap between complex physical properties using General Circulation Models (GCMs), as demonstrated by \citet{selsis_thermal_2011}, and the simplified $f$-factor, shedding light on the importance of atmospheres on rocky tidally-locked exoplanets and providing analytic scalings for how atmospheres impact the dayside brightness temperature. For instance, a thick atmosphere with an $f$-factor of 1/4 can redistribute heat from the dayside to the nightside more efficiently, leading to a lower dayside temperature. In contrast, a no-atmosphere scenario with an $f$-factor of 2/3 cannot transfer any heat from the dayside to the nightside (in a tidally-locked planet), maximizing the dayside temperature ($T_{\rm max} \approx 1.28 T_{\rm eq}$).

Additionally, \citet{mansfield_identifying_2019} proposed a technique to identify the presence of atmospheres on tidally-locked rocky planets by inferring high Bond albedo through cool dayside emission. Observing a high albedo likely indicates the presence of high-albedo clouds, particularly in environments where the atmosphere is not thick enough to redistribute heat but is sufficient for cloud formation, reminiscent of Mars-like conditions. They predicted observed albedo values for various rocky surface compositions (metal-rich, Fe-oxidized, basaltic, ultramafic, ice-rich, feldspathic, granitoid, and clay). All plausible compositions for rocky planet surfaces at 410--1250\,K had predicted albedos $\alpha_{\rm B} < 0.4$ (calculated for TRAPPIST-1b), suggesting that inferred albedos higher than this limit would likely indicate reflective clouds in an atmosphere. 

With these techniques and the emergence of the James Webb Space Telescope (JWST) observations, several studies have aimed to investigate atmospheres on rocky planets around M dwarfs. To date, rocky planet transmission spectra show flat spectra, with no features detected, possibly due to clouds/hazes, high mean molecular weights, or the lack of an atmosphere \citep{lustig-yaeger_jwst_2023,moran_high_2023,may_double_2023,lim_atmospheric_2023,kirk_jwstnircam_2024}. Rocky planet emission spectra have so far been largely consistent with tenuous atmospheres or bare-rock models. For example, \citet{zhang_gj_2024} used JWST's Mid-Infrared Instrument \citep[MIRI;][]{kendrew_mid-infrared_2015} Low Resolution Spectrometer (LRS) mode to probe the hot rocky planet GJ 367b with $T_{\rm eq}$ = 1370\,K by observing the planet's phase curve. They found that the data were consistent with a Planck spectrum with no heat redistribution ($f$-factor = 2/3) and low albedo ($\alpha_{\rm B} \sim 0.1$). TRAPPIST-1b eclipse observations \citep{greene_thermal_2023,ih_constraining_2023} also showed no sign of thick CO$_2$ atmospheres and agreed with bare-rock models when observed using MIRI's F1500W filter (15\,$\mu$m). \citet{xue_jwst_2024} used MIRI/LRS to investigate GJ 1132b and obtain dayside brightness temperature of 709 $\pm$ 31\,K, 1$\sigma$ lower than $T_{\rm max}$ of $746^{+14}_{-11}$\,K consistent with no significant atmosphere both from an energy balance and forward model perspective. Previous works by \citet{kreidberg_absence_2019} and \citet{crossfield_gj_2022} to study the atmospheric retention of rocky exoplanets found no significant atmosphere on either LHS 3844b or GJ 1252b, respectively, using Spitzer/IRAC's channel 2 (4.5 $\mu$m). However, both TRAPPIST-1c \citep[MIRI/F1500W,][]{zieba_no_2023} and GJ 486b \citep[MIRI/LRS,][]{mansfield_no_2024} eclipse depth observations returned slightly less than the instant reradiation expectation, which might hint at a thin atmosphere or non-zero albedo. Interestingly, \citet{hu_secondary_2024} reported a possible CO/CO$_2$ rich atmosphere around 55 Cnc e, a super-hot rocky planet with $T_{\rm eq} \sim$ 2000\,K  using data from MIRI/LRS and NIRCam/grism. This result is similar to the independent study by \citet{patel_jwst_2024}, which used JWST/NIRCam and found the same CO/CO$_2$ rich atmosphere likely outgassed from a magma ocean.

Here, we observe the rocky exoplanet LTT 1445A b with JWST MIRI/LRS to infer its thermal emission spectrum and whether it has a thick atmosphere. At a distance of 6.8638 $\pm$ 0.0012 parsecs \citep{gaia_collaboration_gaia_2023}, LTT 1445A (R = 0.271$^{+0.019}_{-0.010}\,R_{\odot}$ with $T_{\ast}$ = 3340 $\pm$ 150\,K) is the closest known M dwarf to host transiting rocky exoplanets\footnote{As of 4 October 2024, \url{https://exoplanetarchive.ipac.caltech.edu}}. It is the primary star in a triple star system, with lower-mass M dwarf components B and C in a close pair about 7" away. LTT 1445A b is one of two known transiting terrestrial planets in the system, with a radius of 1.34 $^{+0.11}_{-0.06}\,R_{\oplus}$, a mass of 2.73 $^{+0.25}_{-0.23}\,M_{\oplus}$ \citep{pass_hstwfc3_2023}, and a bulk density consistent with a rocky composition at $6.2^{+1.2}_{-1.3}\,{\rm g\,cm^{-3}}$.

LTT 1445A b receives an instellation at 5.7$^{+1.3}_{-1.1}\,S_{\oplus}$ for an equilibrium temperature ($T_{\rm eq}$ = 431 $\pm$ 23\,K) comparable to Mercury ($\sim 6.7\,S_{\oplus}$; $T_{\rm eq} = 439.6$\,K). Previous transmission spectra observed in \citet{diamond-lowe_ground-based_2023} showed a flat line indicating either no atmosphere, an atmosphere with a high mean-molecular-weight (high-MMW), or a hazy/cloudy atmosphere; those transmission spectra rule out a clear, low-MMW atmosphere.

This paper is organized as follows. Section \ref{sec:obs} discusses the details of the JWST observations. ection \ref{sec:methods} describes methods, including the LTT 1445A b system's parameters in Section \ref{subsec:parameters}, JWST data reduction pipelines (Section \ref{subsec:pipelines}), data curation (Section \ref{subsec:trimming}), the LTT 1445A stellar spectrum (Section \ref{subsec:stellar_spec}), and light curve fitting (Section \ref{subsec:lc_fitting}). The results are presented in Section \ref{sec:results}, including the detection of the LTT 1445A b secondary eclipse (Section \ref{subsec:detection_eclipse}), constraints on $e\cos\omega$ and $e\sin\omega$ (Section \ref{subsec:ecosw_esinw}), and the planet's emission spectra (Section \ref{subsec:emission_spectra}). Interpretation and discussion of the results are provided in Section \ref{sec:discussions}, where we examine eccentricity constraints from our observations (Section \ref{subsec:eccentricity}), discuss the planet's dayside brightness temperature and its interpretation (Section \ref{subsec:dayside_temp}), compare the emission spectrum to atmosphere models (Section \ref{subsec:forward_model}), and place LTT 1445A b in context with other planets (Section \ref{subsec:cosmic_shoreline}). We conclude in Section \ref{sec:conclusion}.

\section{Observations} \label{sec:obs}%0.5-1 pages

We observed three eclipses of LTT 1445A b passing behind its star with the Low Resolution Spectrometer (LRS) slitless mode in the Mid-Infrared Instrument (MIRI) on board JWST with the Cycle 1 General Observers (GO) program 2708 (PI: Zach Berta-Thompson). LRS covers wavelengths ($\lambda$) ranging from 5 to 12\,$\mu$m with a spectral resolution that varies from R $\sim$ 40 at $\lambda$ = 5\,$\mu$m to R $\sim$ 160 at $\lambda$ = 10\,$\mu$m \citep{kendrew_mid-infrared_2015}. This wavelength range encompasses the $\lambda$ = 6.7\,$\mu$m predicted spectral radiance peak of the planet, assuming a pure thermal emission from a planet at an equilibrium temperature of 431 $\pm$ 23\,K \citep{pass_hstwfc3_2023}. We observed 5 groups per integration, with a cadence of 0.954 second per integration. We avoided saturation from the relatively bright host star ($J$ = 7.29), with the brightest pixel of the last group reaching 44400 DN (equivalent to 82\% of saturation). We implemented a position angle constraint to keep the B+C components of the LTT 1445 system from overlapping with A in the dispersion direction; no light from these stars was seen on the detector subarray.

The first of the three visits spanned 6.06 hours on-source on 27 August 2022 (16:01:50 - 23:25:34 UT) with a total of 22862 integrations, while the second and third visits lasted 3.76 hours each (23 December 2022, 14:39:33 - 19:44:30 UT, and 5 August 2023, 16:21:10 - 21:26:22 UT) equivalent to 14185 integrations each. The extended duration of the first visit allowed for approximately 3.13 hours before and 1.56 hours after the expected mid-eclipse time, assuming an eccentricity ($e$) of 0, relative to the predicted eclipse duration of 1.38 hours \citep{winters_three_2019}. For the second and third visits, we allowed 1.93 hours before and 0.46 hours after the expected mid-eclipse time. The additional out-of-eclipse time from the first visit helped with our understanding of charge-trapping events at the beginning of observations and characterizing other possible systematics. Moreover, this extra time served as a buffer in case the eclipse occurred outside our original estimated eclipse time assuming $e$ = 0, enabling us to be able to reschedule our second and third visits to better capture the eclipse if necessary.

\section{Methods} \label{sec:methods} % 3 pages
In this section, we will delve into details of our approach to obtain LTT 1445A b's emission spectrum, including our choice of system parameters (Section \ref{subsec:parameters}), data reduction pipelines (Section \ref{subsec:pipelines}), our approach to curate the data (Section \ref{subsec:trimming}), LTT 1445A's instrinsic stellar spectrum (Section \ref{subsec:stellar_spec}), and light curve fitting (Section \ref{subsec:lc_fitting}).

  \subsection{System parameters} \label{subsec:parameters}
The secondary eclipse observations of LTT 1445A b are most sensitive to the planet-to-star flux ratio ($F_p/F_{\ast}$) through the eclipse depth, where the planet flux at MIRI wavelengths is dominated by thermal emission, and to the eccentricity $e$ and argument of periastron $\omega$, through the eclipse timing and duration. The information on other stellar and planetary parameters is minimal. In this work, we adopted stellar and planetary parameters derived from \citet{pass_hstwfc3_2023}, which are the most recent global analysis of the system, shown in Table \ref{tab:parameters}. \citet{pass_hstwfc3_2023} obtained new transit data for planet c from Hubble Space Telescope (HST) using WFC3/UVIS combined with existing data from the Transiting Exoplanet Survey Satellite (TESS), radial velocity (RV) data from ESPRESSO, HARPS, HIRES, MAROON-X, PFS, plus additional 85 ESPRESSO RVs from \citet{lavie_planetary_2023} to help constrain parameters, making this the most comprehensive fit to the system published to date. A circular orbit ($e = 0$) is assumed in this joint analysis, and we verify in this work that the eccentricity of planet b is indeed small. \citet{pass_hstwfc3_2023} did not rederive an effective temperature for the star, instead using $T_{\rm \ast, eff}$ = 3340 $\pm$ 150\,K, from \citet{winters_second_2022}, which included an extra-cautious uncertainty to account for systematic uncertainties in M dwarf bolometric corrections. To get a precise timing of the eclipse, we estimated light travel time delay in the system to be $2a/c$ = 38 s, where $a$ is semi-major axis and $c$ is speed of light. Then, we added this value to the quoted $T_C$ value (transit midpoint) to account for the delay between the position of the planet at transit to its position at eclipse.

\begin{deluxetable*}{cccc}
\tablecaption{Median and 68\% confidence intervals of stellar and planetary parameters used in this work. Most values are updated in \citet{pass_hstwfc3_2023} from \citet{winters_second_2022} by adding HST WFC3/UVIS data and redoing a joint fit between new HST data and TESS data.}
\tablenum{1}
\label{tab:parameters}

\tablehead{\colhead{Parameters} & \colhead{Units} & \colhead{Values} & \colhead{Reference}} 
\startdata
Stellar Parameters: &&&\\
$M_\ast$ & Mass ($M_\odot$)& 0.257 $\pm$ 0.014 & \citet{pass_hstwfc3_2023}\\
$R_\ast$ & Radius ($R_\odot$)& 0.271$^{+0.019}_{-0.010}$& \citet{pass_hstwfc3_2023} \\
$\rho_\ast$ & Density (cgs)& 18.2$^{+2.2}_{-4.0}$ & calculated from \citet{pass_hstwfc3_2023} values \\
$\log g$ & Surface gravity (cgs)& 4.982$^{+0.040}_{-0.065}$& calculated from \citet{pass_hstwfc3_2023} values\\
$T_{\rm \ast, eff}$ & Effective Temperature (K)& 3340 $\pm$ 150& \citet{winters_second_2022}\\
$\left[\text{Fe/H}\right]$ & Metallicity (dex)& -0.34 $\pm$ 0.09 & \citet{winters_second_2022}\\
$d$ & Distance (pc) & 6.8638 $\pm$ 0.0012  & Gaia DR 3 \citep{gaia_collaboration_gaia_2023,gaia_collaboration_gaia_2016}\\
\hline
Planetary Parameters: &&&\\
$P$ & Period (days)& 5.3587635$^{+0.0000044}_{-0.0000045}$& \citet{pass_hstwfc3_2023} \\
$a$ & Semi-major axis (AU)& 0.03810$^{+0.00067}_{-0.00070}$& \citet{pass_hstwfc3_2023}\\
$R_p/R_{\ast}$ &Planet-to-star radius ratio& 0.0454 $\pm$ 0.0012&\citet{pass_hstwfc3_2023}\\  
$R_p$ & Radius ($R_E$)& 1.34$^{+0.11}_{-0.06}$ &\citet{pass_hstwfc3_2023}\\ 
$M_p$ & Mass ($M_E$)& 2.73$^{+0.25}_{-0.23}$&\citet{pass_hstwfc3_2023}\\
$T_C$ & Time of conjunction (BJD$_{\rm TDB}$) \tablenotemark{a}& 2458412.70954$^{+0.00047}_{-0.00046}$& \citet{pass_hstwfc3_2023}\\
$T_{14}$ & Total transit duration (days)& 0.05691$^{+0.00080}_{-0.00075}$ &\citet{pass_hstwfc3_2023}\\
$i$ & Inclination (Degrees)& 89.53$^{+0.33}_{-0.40}$&\citet{pass_hstwfc3_2023} \\
$e$ & Eccentricity\tablenotemark{b} & $<$ 0.110 &\citet{winters_second_2022} \\
%\omega_{\ast}$ & Argument of Periastron (Degrees)& 123$^{+89}_{-67}$ &\\
$\rho_p$ & Density (cgs) & 6.2$^{+1.2}_{-1.3}$ & \citet{pass_hstwfc3_2023} \\
$T_{\rm eq}$ & Equilibrium temperature\tablenotemark{c} (K)& 431 $\pm$ 23 &\citet{pass_hstwfc3_2023}\\
$S$ & Instellation ($S_{\oplus}$) & 5.7$^{+1.3}_{-1.1}$ &\citet{pass_hstwfc3_2023} \\
\enddata
\tablenotetext{a}{We show $T_C$ as measured by a clock located at LTT 1445A b's position at the time of eclipse. We calculate this light travel time corrected $T_C$ by adding $2a/c$ = 38 s to the transit-referenced $T_C$ = 2458412.70910 quoted in \citet{pass_hstwfc3_2023}.}
\tablenotetext{b}{2$\sigma$ (95\%) upper limit. Note that \citet{pass_hstwfc3_2023} assumed $e$ = 0.}
\tablenotetext{c}{Assumes no albedo and perfect heat redistribution}
\end{deluxetable*}

  \subsection{JWST data reduction pipelines} \label{subsec:pipelines}
  We used two data reduction pipelines to ensure the robustness of final spectroscopic light curves: \texttt{Eureka!} \citep{bell_eureka_2022}\footnote{\url{https://github.com/kevin218/Eureka}} and Simple Planetary Atmosphere Reduction Tool for Anyone (\texttt{SPARTA}; first described in \citealt{kempton_reflective_2023})\footnote{\url{https://github.com/ideasrule/sparta}}. These are both open-source packages specifically developed for JWST time-series observations.   
  
    \subsubsection{\texttt{Eureka!}}\label{subsubsec:eureka}
        We ran \texttt{Eureka!} \citep{bell_eureka_2022} version 0.11.dev286+gde5b373b, which uses \texttt{jwst} pipeline version 1.14.0 and \texttt{CRDS} version 11.17.22 for Stage 1 (detector processing) and Stage 2 (data calibration). We then performed spectral extraction and background subtraction using \texttt{Eureka!} (Stage 3). 

        The Stage 1 transition from \texttt{uncal} groups into rates was executed using recommended setup parameters for MIRI/LRS data, with two customizations. First, we turned on \texttt{skip\_firstframe} and \texttt{skip\_lastframe} which will skip \texttt{firstframe} and \texttt{lastframe} correction steps in \texttt{jwst} pipeline, thus preventing the pipeline from flagging and ignoring the first and last groups of each integration. Second, we skipped the \texttt{jwst.jump.JumpStep}, outlier detection meant to correct cosmic rays, since the integrations are short and cosmic rays are rare.
        
  In Stage 2, we performed data calibration steps of flat fielding and unit conversion. We kept Stage 2 control files unchanged from suggested configuration for MIRI/LRS except we skipped the production of \texttt{x1dints} file to speed up the process. 
  
  In Stage 3, we performed background subtraction and spectral extraction with the following changes relative to defaults. We modified the \texttt{gain} parameter, setting it to 3.1 e$^{-}$/ADU instead of the \texttt{jwst} pipeline default of 5.5 e$^{-}$/ADU, in accordance with recommendations outlined in \citep{bell_first_2023}. We conducted several experiments varying the background exclusion half-width (\texttt{bg\_hw}) and spectral aperture half-width (\texttt{spec\_hw}) from \texttt{bg\_hw} = 5 to 10 and \texttt{spec\_hw} = 4 to 8 (always setting the background width larger than aperture width). However, we observed only marginal differences between each pair of values with the median absolute deviation (MAD) of the spectroscopic light curve (Stage 3 product) varying by $\sim$ 1.5\%, and not significantly impacted the final results. Ultimately, we selected \texttt{bg\_hw} = 9 and \texttt{spec\_hw} = 4 since it provided lowest median absolute deviation (MAD). The two-iteration background outlier is extended to full frame with threshold of [5,5] as suggested for shallow transits/eclipses.
  
    \subsubsection{\texttt{SPARTA}} \label{subsubsec:sparta}
        \texttt{SPARTA} is a self-contained package that does not rely on any other pipeline and is widely used in JWST observations \citep[e.g.,][]{kempton_reflective_2023,zhang_gj_2024,xue_jwst_2024,mansfield_no_2024}. Starting from the \texttt{uncal.fits} file, we apply nonlinearity correction, dark subtraction, gain multiplication (assuming a gain of 3.1 e$^{-}$/ADU), up-the-ramp fitting with two rounds of outlier rejection, and flat-fielding. Background subtraction is then performed using the median value of columns [10,21] and [-21,-10] (following Python indexing) for each integration. Finally, optimal extraction is computed by creating a median image and position offset, which are used to calculate a profile. This step employs a half-width window of 6 pixels and excludes pixels that deviate more than 5$\sigma$ from the model as outliers.

\subsection{Data curation} \label{subsec:trimming}
\begin{figure*}
\plotone{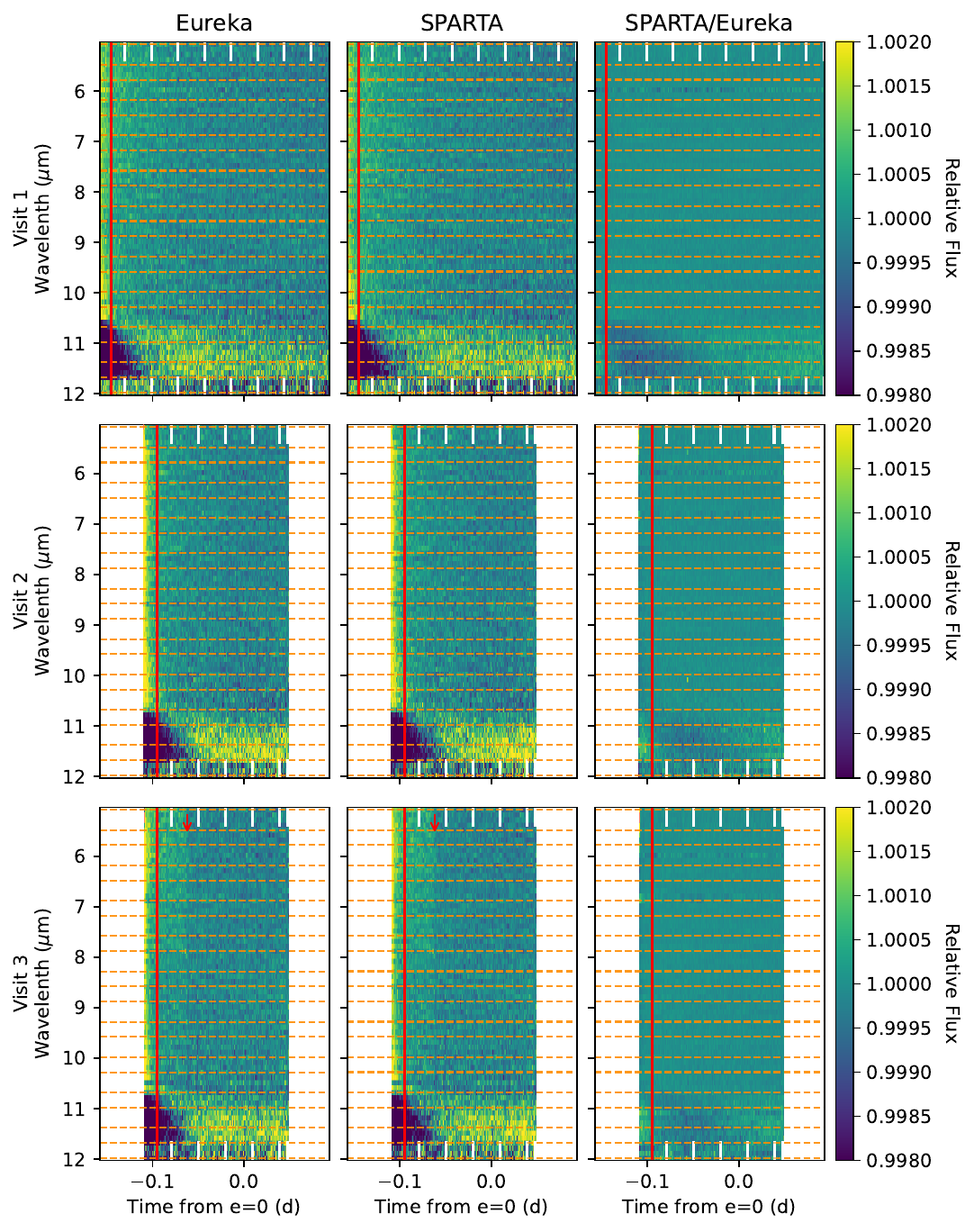}

\caption{Spectroscopic light curves from first, second and third visits (top to bottom) from the \texttt{Eureka!} (left panel) and \texttt{SPARTA} (middle panel) pipelines binned to wavelength resolution of 0.1 $\mu$m and time resolution of 1 minute for better visualization. The ratio between the two pipeline products (SPARTA/Eureka) is shown in the right panel. Each white tick at the top and bottom of each plot indicates the end of each segment of data. Solid vertical red lines indicate 20 minutes of trimming applied to all visits. In visit 3 a clear flux offset is seen near -0.06 days (red arrow), so we trim an addition 50 minutes to avoid it.  Dashed horizontal orange lines indicate the edges of each wavelength bin in 20 bins scheme. The planet's eclipse is too shallow to see with this colorbar.}
\label{fig:raw_data_with_diff}
\end{figure*}

We imported the optimal extracted spectroscopic light curves from \texttt{Eureka!} (optimal extraction) and \texttt{SPARTA}, using \texttt{chromatic}\footnote{\url{https://github.com/zkbt/chromatic/}}, a Python-based tool for reading and visualizing JWST data. We then truncated the \texttt{Eureka!} products below 5 and above 12 $\mu$m to match the wavelength range of data products from \texttt{SPARTA}. Next, we normalized the data by dividing by the median spectrum across all integrations. We flagged outliers more than 10$\sigma$ away from each wavelength's median-filtered light curve and excluded them from future binning or analysis, removing 16 outlying wavelength/time pairs for \texttt{SPARTA} and 71 for \texttt{Eureka!}. The processed data are shown in Figure \ref{fig:raw_data_with_diff}, averaged with inverse-variance weighting into $\Delta \lambda$ = 0.1 $\mu$m wavelength bins and $\Delta t = 60$ s time bins, for visual clarity.

The most notable systematic is an exponential ramp at the beginning of each observation. Such exponential ramps are common in infrared instruments including Spitzer/IRAC \citep{agol_climate_2010}, HST/WFC3 \citep{berta_flat_2012,zhou_physical_2017}, and JWST MIRI/LRS \citep{bell_first_2023,bouwman_spectroscopic_2023} and have been attributed to charge-trapping. However, beyond 10.6 $\mu$m, we observed a sudden change in ramp behavior, from a downward lower amplitude ramp to an upward higher amplitude ramp (see Figure \ref{fig:raw_data_with_diff}). This is consistent with the ``shadowed'' region described by \citep{bell_first_2023}, where pixels on either side of this wavelength may experience qualitatively different illumination histories. This phenomenon appears to be common but not universal across MIRI/LRS observations, seen in \citet{bell_nightside_2024, kempton_reflective_2023, zhang_gj_2024, hu_secondary_2024,mansfield_no_2024} but not in \citet{welbanks_high_2024,xue_jwst_2024}. 

To mitigate the charge-trapping ramp, we removed the first 20 minutes of data from all three visits (solid red vertical lines in Figure \ref{fig:raw_data_with_diff}), during which this effect was strongest. While this cutoff does not completely eliminate the charge-trapping ramp, especially in the shadowed region where the direction of the ramp changes from fading to brightening, it significantly simplifies the systematic so that it can be well-modeled with a single exponential function during light curve fitting (see Section \ref{subsec:lc_fitting}). During the third visit, we observed a sudden drop in flux in both \texttt{Eureka!} and \texttt{SPARTA} products (see Figure \ref{fig:raw_data_with_diff} bottom panel), at around the 350 ppm level spanning approximately 9 minutes (broadband light curve from 5-12\,$\mu$m). This drop occurred at BJD time = 2460162.280 (about 1 hour from the start of the observation). This drop is not at the gap of a data sector nor is it likely to be our target's secondary eclipse (the calculated eclipse depth is 28\,ppm at 8.54\,$\mu$m). This flux drop is similar to one seen the MIRI/LRS observation of GJ 367b in \citet{zhang_gj_2024}; it is not immediately clear whether its cause is a detector state change or something else. We did not find any correlation associate with FGS data using \texttt{Spelunker} \citep{deal_spelunker_2024} so we think that this is not a mirror-tilt event. We mitigate this sudden drop by trimming off an additional 50 minutes (a total of 70 minutes) for the third visit.

  \subsection{LTT 1445A stellar spectrum} \label{subsec:stellar_spec}
\begin{figure}

\plotone{ 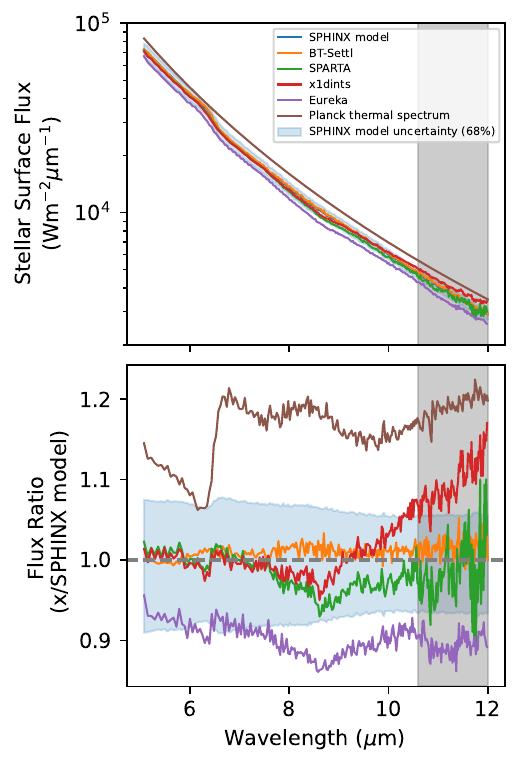}
\caption{Possible stellar spectra of LTT 1445A, including our adopted SPHINX model and other theoretical or empirical options. Spectra are shown both as surface flux (top) and as a ratio to the adopted SPHINX spectrum (bottom). The blue shading indicates the effect on the spectra of changing stellar effective temperature by $\pm$ 150\,K (1$\sigma$). Wavelengths beyond 10.6 $\mu$m (shaded in gray) fall within the shadowed region and may have less reliable observed absolute fluxes.}
\label{fig:stellar_compare}
\end{figure}

It is crucial to determine the host star spectrum at MIRI/LRS wavelengths to be able to compare the planet's thermal emission to theoretical models. The measured secondary eclipse depth is a planet-to-star relative flux contrast ($D_{sec} = F_p/F_{\ast}$), and it needs to be multiplied by a stellar spectrum to convert to the planet's intrinsic dayside thermal emission flux. In this section, we will explore different options for stellar spectra of LTT 1445A and estimate the uncertainty on our adopted spectrum. Figure \ref{fig:stellar_compare} shows LTT 1445A spectra from different sources as described below. In all cases, the stellar flux is represented at the surface of the star, as the $\mathrm{W}\mathrm{m}^{-2}\mu \mathrm{m}^{-1}$ that could be multiplied by the stellar surface area $4\pi R_\ast^2$ to get to the star's luminosity. 

\begin{itemize}
    
        \item
        The SPHINX model: \citet{iyer_sphinx_2023} designed this grid to improve the treatment of broadband molecular features in M dwarf stars and benchmarked it against empirical spectra. We interpolated the pre-calculated grid spectra to the parameters in Table \ref{tab:parameters} assuming solar C/O ratio of 0.5. We performed trilinear interpolation in log-surface gravity ($\log g$), log-metallicity ($\log_{10} Z$), and log stellar effective temperature ($\log T_{\rm \ast, eff}$), since fluxes per wavelength can grow $\propto T_{\rm \ast, eff}^x$ where $x \neq 1$. We identified stellar effective temperature ($T_{\rm \ast, eff}$ = 3340 $\pm$ 150\,K, 4.5\% uncertainty, \citealt{winters_second_2022}) as the main source of uncertainty in our stellar spectra. We perturbed $T_{\rm \ast, eff}$ by $\pm$ 150\,K ($T_{\rm \ast, eff}$ = 3190, 3490\,K) resulting in spectra that are, on average, 6\% higher and 7\% lower (shown by the blue shading in Figure \ref{fig:stellar_compare}). We adopt this model and uncertainty as our reference stellar spectrum for the rest of this work.
        
        \item BT-Settl model:
        \citet{diamond-lowe_high-energy_2024} published a panchromatic spectral energy distribution (SED)  comprising empirical data in X-ray, UV, and blue optical, estimates of EUV, Ly$\alpha$ line reconstruction, and BT-Settl (CIFIST) model for the rest of optical and infrared. In the MIRI/LRS wavelength range the spectrum is just the BT-Settl (CIFIST) model \citep{allard_model_2003,caffau_solar_2011} using stellar parameters as in \citet{winters_second_2022}. \citet{diamond-lowe_high-energy_2024} verified that the normalization of this spectrum closely matches the observe low-resolution prism spectrum of LTT 1445A from Gaia. This empirically-informed BT-Settl model is higher than one produced by SPHINX model by about 1\% and well within our adopted SPHINX uncertainty.
        
        \item \texttt{Eureka!} MIRI/LRS spectrum:
        We derived flux-calibrated stellar spectra from Visit 1 with \texttt{Eureka!} by turning on \texttt{calibrated\_spectra}, and with guidance and caution about the reliability of this process from \texttt{Eureka!} developers (Taylor Bell, private communication). To better capture the full stellar flux for the intrinsic stellar spectrum, we used a box extraction with spectral aperture (\texttt{spec\_hw}) and background half-width (\texttt{bg\_hw}) of 9 and 15 pixels respectively, larger than the 4 and 9 pixels used for time-series extraction (see Section \ref{subsubsec:eureka}). We then performed an aperture correction to translate from finite spectral aperture to infinite aperture using JWST's \texttt{jwst\_miri\_apcorr\_0012.fits} file and translated the observed flux at JWST into a stellar surface flux in $\mathrm{W}\mathrm{m}^{-2}\mu \mathrm{m}^{-1}$ assuming the distance and radius reported in Table \ref{tab:parameters}. When compared to the SPHINX model, the time-integrated observed flux from \texttt{Eureka!} at \texttt{spec\_hw} = 9 is $\sim$10\% lower than the $T_{\rm \ast, eff}$ = 3340\,K SPHINX model on average, about 1-2 $\sigma$ given the estimated SPHINX uncertainty. 
        
        %Even though we performed aperture correction on both \texttt{Eureka!} products (\texttt{spec\_hw} of 4 and 9) but due to systematics and possible inconsistency between \texttt{Eureka!} pipeline and \texttt{jwst} pipeline. This drives the difference between \texttt{spec\_hw} of 4 and 9.
        
        %Even though, \texttt{spec\_hw} = 4 pixels provided spectrum that on average is 3 \% lower than the SPHINX model which at first seems better than \texttt{spec\_hw} = 9. Absolute fluxes from aperture half-width of 4 did not agreed with aperture half-width of 20 (which aperture correction is almost 1.0 for all wavelengths) and found that using the smaller aperture optimized for light curve precision resulted in unreliable estimate of the absolute stellar flux. 

        \item \texttt{SPARTA} MIRI/LRS spectrum:
        We performed a similar reduction to that described in Section \ref{subsubsec:sparta}; however, using a simpler box extraction with an aperture half-width of 20 pixels. Using a larger aperture window ensures that we have captured almost all the flux and that the aperture correction array is almost 1.0 across the wavelength range (max correction is 1.0035). The SPARTA spectrum agrees closely with the adopted SPHINX spectrum, although with systematic offsets up to 5\% at longer wavelengths.
        
        \item STScI pipeline (x1dints) MIRI/LRS spectra:
        We downloaded the Visit 1 flux-calibrated JWST official \texttt{x1dints} file from the MAST portal and compared it to other spectra. We found mostly agreement with other spectra below 10 $\mu$m, however, as this default pipeline product does not include background subtraction it may be increasingly subject to telescope thermal radiation at longer wavelengths.

        \item Planck thermal spectrum:
        For reference, we show a simple Planck thermal emission spectrum with $T_{\rm \ast, eff}$ = 3340 K. It overestimates the stellar flux by 10-20\% and entirely misses the broad jump at 6-7 $\mu$m present in the other theoretical and empirical spectra (see figure \ref{fig:stellar_compare}, bottom panel).
\end{itemize}
  
  \subsection{Light curve Fitting} \label{subsec:lc_fitting}
We performed light curve fitting using \texttt{chromatic\_fitting} (Murray et al., in prep.), a Python-based open-source tool specifically designed for multiwavelength light curve fitting. This tool uses \texttt{PyMC3} for efficient inference of parameter posterior probability distributions \citep{salvatier_probabilistic_2016}. For all eclipse fits in this paper, we use \texttt{EclipseModel}, which is a wrapper of the \texttt{starry} package \citep{luger_starry_2019}.

For non-zero eccentricities ($e > 0$), directly including the argument of periastron $\omega$ as a model parameter can be challenging due to its degeneracy with $\omega+\pi$. One approach is to re-parameterize in $e\cos\omega$ and $e\sin\omega$, which ties strongly with observable eclipse timing and duration, respectively \citep{winn_transits_2014}. However, sampling in $e\cos\omega$ and $e\sin\omega$ induces a linear prior in $e$ when applying a uniform prior in $e\cos\omega$ and $e\sin\omega$ \citep[see detailed discussions in][]{ford_improving_2006, eastman_exofast_2013}. Instead, we opted to re-parameterize in $\sqrt{e}\cos\omega$ and $\sqrt{e}\sin\omega$, as suggested by \citet{anderson_wasp-30b_2011}, which simplifies model ambiguities associated with the $\omega$ parameter while correcting for a uniform prior in $e$.

After carefully examining the data, we employed the following combination of systematic and eclipse models:
\begin{eqnarray}
    F_{\rm obs}(t,\lambda) &=& F_{\rm eclipse}(t,\lambda) \times F_{\rm systematics}(t,\lambda).\label{eqn:fitting_components}
\end{eqnarray}

Here $F_{\rm eclipse}(t,\lambda)$ represents the flux computed from the \texttt{EclipseModel} within \texttt{chromatic\_fitting}. $F_{\rm systematics}(t,\lambda)$ is the systematics model built from a combination of \texttt{PolynomialModel} and \texttt{ExponentialModel} models in \texttt{chromatic\_fitting}:
\begin{eqnarray}
    F_{\rm systematics}(t,\lambda) &=& F_{\rm time}(t,\lambda) F_{\rm ramp}(t,\lambda) \label{eqn:systematics_components} \\ &&F_y(t,\lambda) F_{x}(t,\lambda). \nonumber
\end{eqnarray}

$F_{\rm time} = p_{t,0} + p_{t,1}t_{\rm rel}$ is a first-order polynomial in relative time, $t_{\rm rel} = t-\Bar{t}$ (where $\Bar{t}$ is the mean time of the trimmed data), used to correct for any linear trends and the baseline stellar flux. $F_{\rm ramp} = 1 + A\exp\left[\frac{-(t_{\rm rel} - t_{\rm rel,0})}{\tau}\right]$ is an exponential ramp in relative time aimed at describing the residual charge-trapping ramp common in mid-infrared instruments \citep[e.g.,][]{agol_climate_2010,berta_flat_2012,zhou_physical_2017,bell_first_2023} while $t_{\rm rel,0}$ is a relative start time of the time series after cutting off the first 20 minutes of the data (70 minutes in visit 3). Notably, the exponential amplitude $A$ can be positive or negative to account for the different ramp behavior for wavelength above (downward ramp) and below 10.6 $\mu$m (upward ramp). $F_y$ = 1.0 + $p_{y,1}y(t)$ and $F_{x}$ = 1.0 + $p_{x,1}x(t)$ are linear functions of the y (spectral axis) and x (spatial axis) centroids in each frame, respectively, for data from the \texttt{SPARTA} pipeline. For \texttt{Eureka!} data, instead of the spectral axis (y) centroid, we used the provided spatial width of the PSF ($\sigma_y$). Note that \texttt{Eureka!} rotated each frame by 90 degrees, so the x and y axes were swapped from the convention (and the \texttt{SPARTA} pipeline).

For all fitting, we binned the data to a time increment of 1 minute to improve computational efficiency and account for correlated noise on a shorter timescale. We fixed the stellar radius, stellar mass, orbital period, time of conjunction, orbital inclination, planet mass, and planet radius to the values in Table \ref{tab:parameters}. We fixed the stellar amplitude to 1.0 to migrate any baseline in normalized stellar flux to the $p_{t,0}$ term in $F_{\rm time}$. We also set the stellar and planet brightness maps to be uniform so other \texttt{Starry} parameters will not affect our light curve. The secondary eclipse depth then can be directly interpreted as a planet brightness map overall amplitude, which is propositional to planet's luminosity. The priors of eclipse depth and [$\sqrt{e}\cos\omega, \sqrt{e}\sin\omega$], which ultimately control eclipse timing, will be later discussed in Sections \ref{subsubsec:broadband} and \ref{subsubsec:fixe}.

The \texttt{PyMC3} module \citep{salvatier_probabilistic_2016}, integrated into \texttt{chromatic\_fitting}, was employed for parameter space exploration using No-U-Turn Sampler \citep[NUTS;][]{hoffman_no-u-turn_2014} Markov Chain Monte Carlo (MCMC) techniques with 80,000 burn-in steps and 50,000 subsequent runs utilizing 4 chains. We included an uncertainty inflation ratio ($n_{\rm \sigma,fitting}$) to account for extra scatter in flux. The uncertainty inflation ratio ($n_{\rm \sigma,fitting}$) is defined as the fitted parameter needed to inflate the light curve uncertainty until a reduced $\chi^2$ of 1 is met. Hence, one can use $n_{\rm \sigma,fitting}^2$ to assess the goodness-of-fit similar to reduced $\chi^2$. The Gelman-Rubin diagnostic test \citep{gelman_inference_1992, brooks_general_1998} was used to indicate convergence, with $\hat{R} = \frac{\hat{V}}{W} < 1.001$, where $\hat{V}$ is the variance of the posterior between chains and $W$ is the variance within each chain.

\subsubsection{Broadband fit to orbital parameters} \label{subsubsec:broadband}
We started by fitting a wavelength-integrated broadband light curve to determine best-fit values for $e$ and $\omega$ since these would be the best constraints on planet eccentricity and argument of periastron ($\omega$) over other methods such as RV. Also, we want to establish confidence that we have captured the secondary eclipse of LTT 1445A b. We constructed this light curve by excluding data redder than 10.62 $\mu$m because of the qualitatively different systematics in the shadowed region (Figure \ref{fig:raw_data_with_diff}) and then averaging together the normalized pixel light curves using inverse-variance weighting. We also calculated a flux-inverse-variance weighted average wavelength called the ``effective'' wavelength. This effective wavelength serves as a better representation of characteristic wavelength for each wavelength bin since it is weighted the same way as the fluxes. 

For the eclipse depth in the broadband fit only, we sample in $\log_{10}(\rm eclipse\,depth)$ with a uniform prior = [-6,-3]. This ensures the eclipse depth must be positive, which will help constrain both $e\cos\omega$ and $e\sin\omega$ better. Uniform prior in a log-space also helps favor shallow eclipse depth and better perform a uniform prior in linear space. It is important to note that we do not use the eclipse depth from this step in our final emission spectrum, where the requirement that eclipse depths must be greater than 0 could bias the posteriors.

For the parameters $[\sqrt{e}\cos\omega, \sqrt{e}\sin\omega]$, we employed uniform priors ranging from -0.332 to 0.332. This range corresponds to a maximum value of $e \approx 0.11$, corresponding to the 95\% confidence upper limit derived from radial velocities \citep{winters_second_2022}.

    \subsubsection{Spectrophotometric light curve fitting} \label{subsubsec:fixe}
After obtaining the best-fit $\sqrt{e}\cos\omega$ and $\sqrt{e}\sin\omega$ values from Section \ref{subsubsec:broadband}, we fixed these values and proceeded to fit the eclipse depth in different wavelength bins. In each bin, we fit the eclipse model simultaneously with the systematics model described in Equation \ref{eqn:systematics_components}. We used 1, 5, 10, and 20 wavelength bins (binning edges are shown in Table \ref{tab:depth_sparta}), where light curves in each bin were constructed through inverse-variance weighting of its constituent normalized pixel light curves. We purposely placed bin edges such that the last bin, last two bins, and last four bins in the 5, 10, and 20 bin schemes, respectively, are entirely in the shadowed region. Therefore, we can assess the effect of the sudden change in systematic behavior on our results by simply excluding the shadowed region bins from the analysis.

For the emission spectrum eclipse depths, we used a uniform prior in the range [$-10^{-3}$, $10^{-3}$]. We allowed the eclipse depth to take negative values, which would imply that the system brightens during the eclipse. Although this is not physical, allowing negative eclipse depths facilitates straightforward statistical analysis and helps mitigate the asymmetrical posterior distribution characteristic of low SNR observations.
    
\section{Results} \label{sec:results} % 3-4 pages
In this section, we present the results of our analysis. We first discuss adopting the \texttt{SPARTA} pipeline as our primary data product for further discussion (Section \ref{subsec:adopt_pipeline}). Then, we confirm the detection of the secondary eclipse (Section \ref{subsec:detection_eclipse}), followed by the constraints on $e\cos\omega$ and $e\sin\omega$  (Section \ref{subsec:ecosw_esinw}). Finally, we introduce the emission spectrum of LTT 1445A b along with $\chi^2$ statistic comparisons to key hypothetical dayside temperatures (Section \ref{subsec:emission_spectra}).

\subsection{Adopted Extraction Pipeline}\label{subsec:adopt_pipeline}
\begin{figure}
\plotone{ 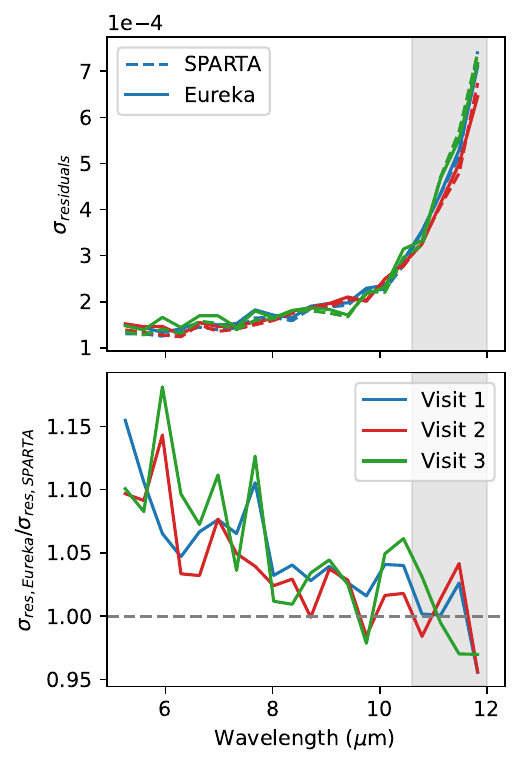}
\caption{({\em Top}) A plot showing the measured scatter in the model residuals with 20 wavelength binning for each of the three visits ($\sigma_{\text{residuals}}$) from the \texttt{Eureka!} (solid line) and \texttt{SPARTA} (dashed line) pipelines. ({\em Bottom}) The ratio of measured scatter from \texttt{Eureka!}/\texttt{SPARTA} ($\sigma_{\rm res,Eureka}$/$\sigma_{\rm res,SPARTA}$). The grey bands show shadowed region at 10.6 to 12 $\mu$m. The measured scatter is calculated as the standard deviation of the 1-minute time bin residuals for each of the 20 wavelength bins described in Section \ref{subsubsec:fixe}.}
\label{fig:plot_noise_comparison}
\end{figure}

Figure \ref{fig:raw_data_with_diff} shows the extracted fluxes from \texttt{SPARTA} and \texttt{Eureka!} are broadly similar, but the most direct quantitative way to compare them is through the residuals to model fits. We performed broadband and spectrophotometric fits for both pipelines' extracted time-series spectra. In the broadband fits (Section \ref{subsubsec:broadband}), the expected per-point uncertainty from \texttt{SPARTA} is 35 ppm at 1 minute cadence, and the achieved scatter of the residuals to the broadband eclipse is 41, 44, 42 ppm for the three visits. For \texttt{Eureka!}, the expected per-point uncertainty at the same cadence is 32 ppm, while the achieved scatter of the residuals is 52, 50, 53 ppm for the three visits, $\sim$20\% higher than with \texttt{SPARTA}. For the spectrophometric fits, Figure \ref{fig:plot_noise_comparison} compares the measured scatter in the model residuals for both extractions, in 20 wavelength bins for the three visits. At the longest wavelengths, where the photon noise is intrinsically higher due to low stellar flux and instrument throughput, the two pipelines closely agree. Toward shorter wavelengths, where the photon noise is lower and subtleties of the extraction matter more, the \texttt{Eureka!} residuals show approximately 10-15\% higher scatter than \texttt{SPARTA}. Given \texttt{SPARTA}'s slight improvement in all fits, we adopt these results for our primary conclusions throughout the rest of the paper. In all analyses we confirmed that \texttt{Eureka!} light curves lead to consistent results, albeit with slightly larger final uncertainties.

\subsection{Eclipse Detection} \label{subsec:detection_eclipse}
Before delving into details about the constraints on eccentricity and the emission spectra, it is crucial to confirm that we have captured the secondary eclipse of LTT 1445A b. The planet's thermal emission is close to the noise level, and given the potential of the inner planet LTT 1445A c (historically discovered after planet b) to excite eccentricity that might move LTT 1445A b's eclipse away from the time predicted for a circular orbit, we first demonstrate that we clearly detect the eclipses in this MIRI/LRS dataset.

\begin{figure*}
\plotone{ 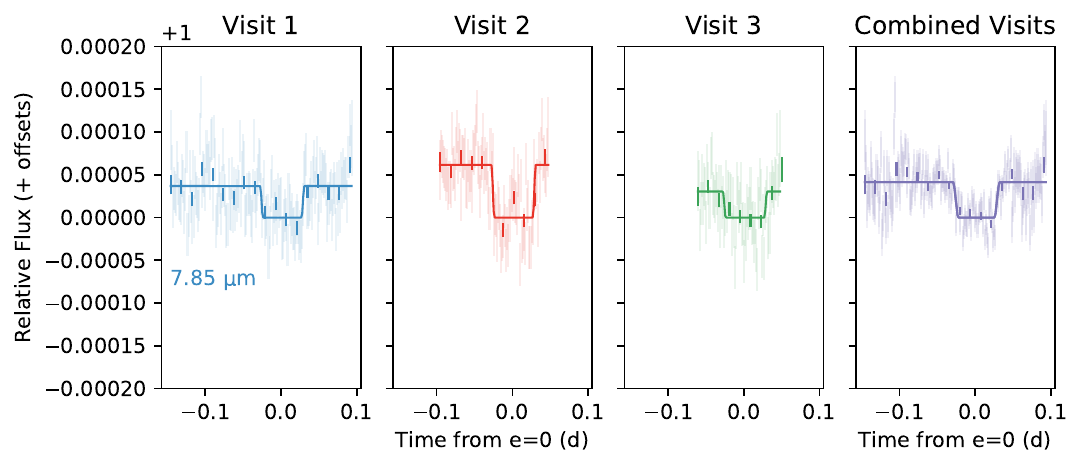}
\caption{Broadband (5--10.62\,$\mu$m) systematics-corrected light curves for three individual visits and a weighted average across all visits. Data are shown from the \texttt{SPARTA} extracted fluxes. Light curves are binned in wavelength in time to $dt = 2$ minutes (transparent error bars) and $dt = 20$ minutes (opaque error bars) for visualization purposes only. Eclipse models are shown in solid lines using depths from the individual-visit and weighted-average values in Table \ref{tab:depth_sparta}.}
\label{fig:lc_systematics_removed_1}
\end{figure*}

\begin{figure*}
\plotone{ 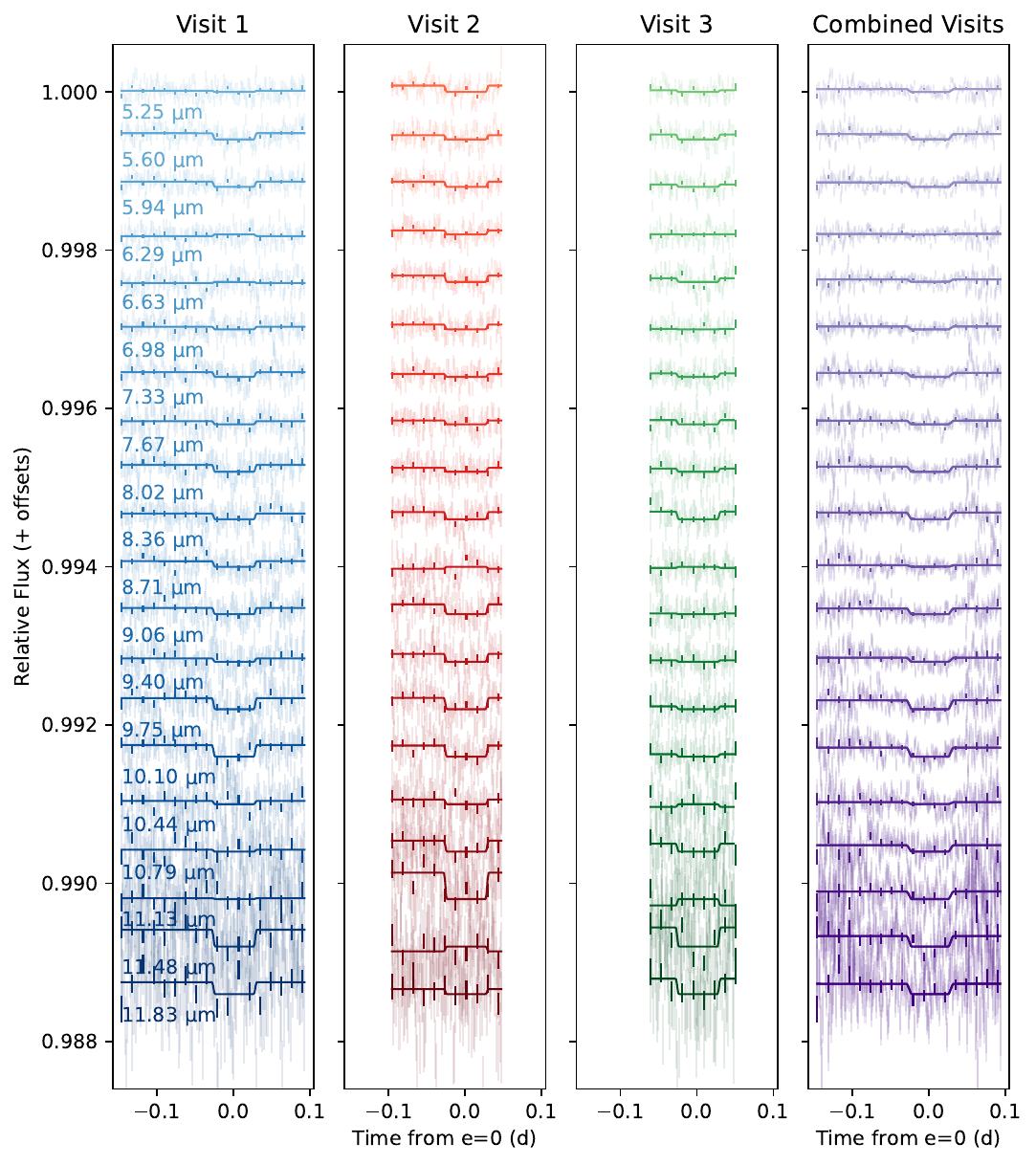}
\caption{Panchromatic systematics-corrected light curves for three individual visits (left to right) and an inverse-variance weighted average across all visits (far-right column). Data are shown from the \texttt{SPARTA} extracted fluxes; \texttt{Eureka!} light curves are  similar. 
Light curves are binned in wavelength to $d\lambda$ = 0.35\,$\mu$m (20 wavelengths) and in time to $dt = 2$ minutes (transparent error bars) and $dt = 20$ minutes (opaque error bars). Eclipse models are shown in solid lines using depths from the individual-visit and weighted-average emission spectra in Table \ref{tab:depth_sparta}.}
\label{fig:lc_systematics_removed_10}
\end{figure*}

\begin{figure*}
\plotone{ 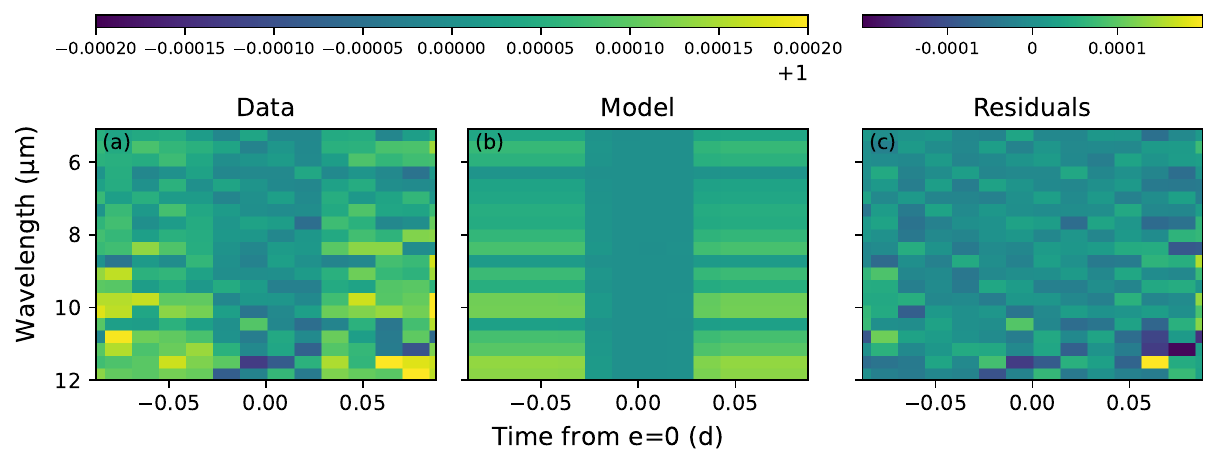}
\caption{Maps of the systematics-corrected and visit-combined flux of the system, binned into large $d\lambda$ = 0.35\,$\mu$m wavelength bins and $dt = 20$ minutes time bins to decrease noise for visual display. Flux is normalized to 1 during eclipse, when the planet is behind the star. Both data ({\em left}) and best-fit eclipse models ({\em middle}) generally show the planet contributing more flux at longer wavelengths; residuals (data - model; {\em right}) are consistent with noise.}
\label{fig:imshow_systematics_removed_10_with_res}
\end{figure*}

Secondary eclipses of LTT 1445A b are not easily visible in the raw data (Figure \ref{fig:raw_data_with_diff}). However, if we use the best-fit (means of posterior distributions) model parameters from a simultaneous eclipse and systematics fit (Equation \ref{eqn:fitting_components}) for each visit, and remove the systematics contribution from the model (Equation \ref{eqn:systematics_components}), the eclipses emerge. Figures \ref{fig:lc_systematics_removed_1} and \ref{fig:lc_systematics_removed_10} show systematics-corrected light curves for the broadband and spectrophotometric fits respectively (20 wavelength bins). We overplot individual visit eclipse model fits using solid lines. We also included the inverse-variance weighted average of the three visits along with an eclipse model generated from the final eclipse depths in the far-right column (see Section \ref{subsec:emission_spectra} and Table \ref{tab:depth_sparta} for the complete set of depths). Figure \ref{fig:imshow_systematics_removed_10_with_res} shows the same systematics-corrected and combined dataset as the last column of Figure \ref{fig:lc_systematics_removed_10}, binned and rendered as 2D flux maps. Eclipses occur with consistent timing and depths across the three independent visits, and the eclipse depth grows toward longer wavelengths, as qualitatively expected for thermal emission at these wavelengths. These two independent pieces of evidence (the eclipse timing and the depth) emphasize that we truly detected the secondary eclipse of LTT 1445A b.

Additionally, the uncertainty inflation ratio ($n_{\sigma,\text{fitting}}$), that is effectively the ratio needed to achieve a corrected reduced chi-squared $\chi^2_{\nu,corr}$ = $\chi^2_{\nu,uncorr}/n_{\sigma,\text{fitting}}^2$ of 1.0, indicates that the secondary eclipse model is slightly preferred. With an eclipse included in the model, we find $n_{\sigma,\text{fitting}}^2 \approx \chi^2_{\nu,uncorr}$ = 1.35 $\pm$ 0.10, 1.50 $\pm$ 0.15, and 1.41 $\pm$ 0.16 for first, second, and third visit; in contrast, if we model the light curves only with systematics, with no eclipse included, we find $n_{\sigma,\text{fitting}}^2$ = 1.46 $\pm$ 0.11, 1.82 $\pm$ 0.18, and 1.51 $\pm$ 0.18 (all from broadband light curves). Note that higher $\chi^2_{\nu,uncorr}$ indicates a worse fit if no eclipse is included.
    
\subsection{$e\cos\omega$ and $e\sin\omega$} \label{subsec:ecosw_esinw}
The observations were planned assuming a near-circular orbit, but we test that assumption by allowing for a non-zero eccentricity that might change the eclipse's timing through $e\cos\omega$ or its duration through $e\sin\omega$. If the eccentricity is large, assuming a perfectly circular orbit could bias eclipse depths by missing the eclipse time or using the wrong eclipse duration. Here we summarize what we learn about $e$ and $\omega$ through fits to the broadband light curves.

\begin{figure}
\plotone{ 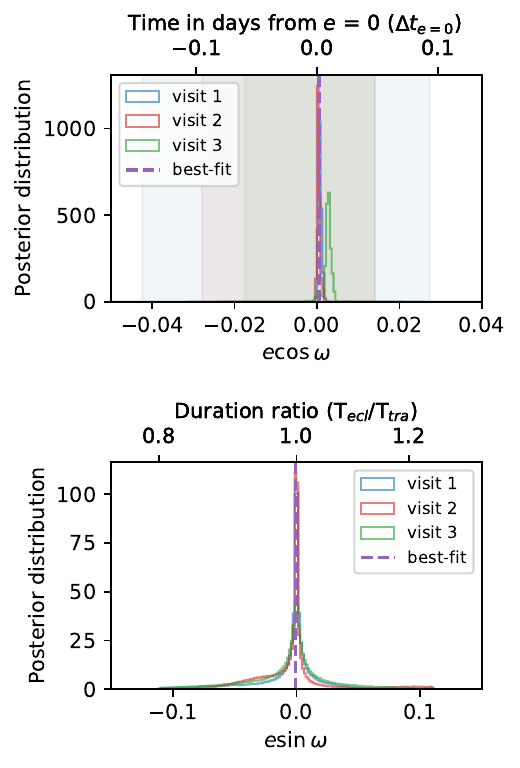}
\caption{Posterior histograms of $e\cos\omega$ ({\em top}) and $e\sin\omega$ ({\em bottom}) along with physical interpretation axes for $e\cos\omega$ which sets secondary eclipse timing (Equation \ref{eqn:ecosw}) and $e\sin\omega$ which sets secondary eclipse duration (Equation \ref{eqn:esinw}). Each vertical colored-region in top panel indicates the observing duration of each visit. The purple dashed lines indicates the adopted values, which are the inverse-variance weighted average of the best fits from the three visits.}
\label{fig:ecosw_esinw}
\end{figure}

Even though we fit for $\sqrt{e}\cos\omega$ and $\sqrt{e}\sin\omega$, the results in this section will be mainly given in $e\cos\omega$ and $e\sin\omega$ since it ties more strongly with observables. We calculated $e\cos\omega$ and $e\sin\omega$ by transforming MCMC samples in $\sqrt{e}\cos\omega$, $\sqrt{e}\sin\omega$ to eccentricity $e$ = $(\sqrt{e}\cos\omega)^2$ + $(\sqrt{e}\sin\omega)^2$ and argument of periastron $\omega$ = $\arctan(\frac{\sqrt{e}\sin\omega}{\sqrt{e}\cos\omega})$. Then, we reconstruct 
$e\cos\omega$ and $e\sin\omega$, showing their posteriors in Figure \ref{fig:ecosw_esinw}. In the limit of small eccentricity, we can add a second x-axis to the top panel of Figure \ref{fig:ecosw_esinw} as the offset of the mid-eclipse time from the $e$ = 0 prediction ($\Delta t_{e = 0}$) using Equation 33 in \citet{winn_transits_2014}:
\begin{eqnarray}
    \Delta t_{e = 0} \approx \frac{2P}{\pi}e\cos\omega ,\label{eqn:ecosw}
\end{eqnarray}
where $P$ is the planet's orbital period and the light-travel-time delay has been accounted for in Table \ref{tab:depth_sparta}. Likewise, for the bottom panel, we can translate $e\sin\omega$ to duration ratio ($T_{\rm ecl}/T_{\rm tra}$) using Equation 34 in \citet{winn_transits_2014}:
\begin{eqnarray}
    \frac{T_{\rm ecl}}{T_{\rm tra}} \approx \frac{1 + e\sin\omega}{1 - e\sin\omega}, \label{eqn:esinw}
\end{eqnarray}
where $T_{\rm tra}$ and $T_{\rm ecl}$ are the transit and secondary eclipse durations respectively. 

We observed good agreement within 1$\sigma$ of $e\cos\omega$ between the first visit, $e\cos\omega$ = 0.00069 $\pm$ 0.00042, and the second visit, $e\cos\omega$ = 0.00031 $\pm$ 0.00037. While the posterior peak in $e\cos\omega$ for visit 3 is higher, at 0.0027 $\pm$ 0.0129, a greater uncertainty is observed due to a shorter out-of-eclipse baseline, even shorter than the in-eclipse duration (note that we also observed a bump in the visit 3 $e\cos\omega$ posterior, overlapping with peaks from the first and second visit). We calculated the weighted average $e\cos\omega$ of all visits as 0.00048 $\pm$ 0.00028, which translates to 2.4 $\pm$ 1.4 minutes later than the $e = 0$ (circular orbit) expectation, shown as the purple dashed line in the top panel of Figure \ref{fig:ecosw_esinw}. The combined $e\cos\omega$ value is consistent with 0 at 1.7$\sigma$.

For $e\sin\omega$, constraints from the eclipse duration are less precise than $e\cos\omega$, as expected, but they have good agreement across the three visits (see Figure \ref{fig:ecosw_esinw}; bottom). The best-fit value for $e\sin\omega$ is $-0.0005 \pm 0.0173$ (purple dashed vertical line in bottom panel), statistically indistinguishable from circular. This value of $e\sin\omega$ also translates to the eclipse-to-transit duration ratio ($T_{\rm ecl}/T_{\rm tra}$) of 1.00 $\pm$ 0.04.

Moreover, constraints on $e\cos\omega$ and $e\sin\omega$ act as an independent check on the presence of eclipses. Given our wide uniform priors on [$\sqrt{e}\cos\omega$, $\sqrt{e}\sin\omega$], the eclipses could theoretically fall outside our observation windows (indicated as colored shade in Figure \ref{fig:ecosw_esinw}). However, the fact that the posteriors find consistent eclipse timings and durations (that are also very close to circular) is another indication that real eclipses are detected.

Although not discussed in detail, \texttt{Eureka!}'s products' best-fit $e\cos\omega$ and $e\sin\omega$ also agreed well within 1$\sigma$ with best fit $e\cos\omega$ = 0.00030 $\pm$ 0.00035, and $e\sin\omega$ = -0.0002 $\pm$ 0.0219 translates to eclipse time delay of 1.5 $\pm$ 1.7 minutes and eclipse-to-transit duration ratio of 1.00 $\pm$ 0.04. 

%We also conducted experiment with different prior on eclipse depth to get the best $e\cos\omega$ and $e\sin\omega$ with uniform prior in linear space from [-10$^{-3}$,10$^{-3}$] (uniform prior and sample in linear space), uniform prior where we sample eclipse depth in log-space from [10$^{-6}$,10$^{-3}$] (loguniform prior and sample), and laplace prior (laplace prior but sample in linear) as stated in Section \ref{subsubsec:broadband}. We found no difference for combined visit 1 and 2 for \texttt{SPARTA} pipeline data and retrieved $e\cos\omega$ and $e\sin\omega$ = 0.00062 $\pm$ 0.00028 and -0.0005 $\pm$ 0.0207 for log-uniform case. For uniform prior, we got -0.00917 $\pm$ 0.00026 and 0.006 $\pm$ 0.021 for $e\cos\omega$ and $e\sin\omega$ only first visit. However, for second visit we got $e\cos\omega$ and $e\sin\omega$ = -0.01543 $\pm$ 0.00033 and 0.015 $\pm$ 0.032 respectively. This is not agreed with log-uniform and Laplace priors. But the best-fit eclipse depth is -78 $\pm$ 13 ppm which is not physical, we disregard this result since the model fits for un-physical noise and skewed the posterior of $e\cos\omega$ and $e\sin\omega$.

 \subsection{Emission Spectra} \label{subsec:emission_spectra}
\begin{figure*}
\plotone{ 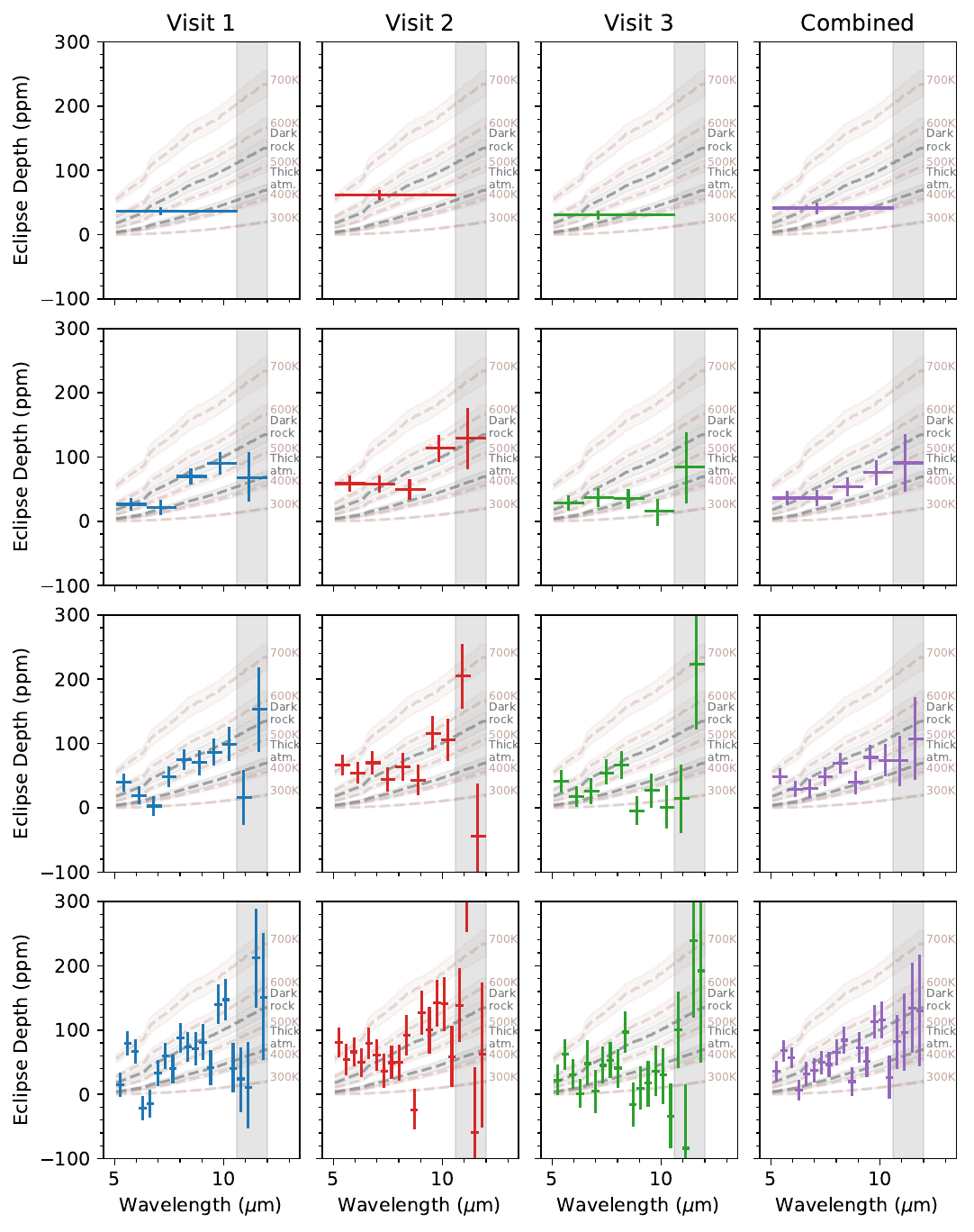}
\caption{Emission spectra in terms of measured secondary eclipse depth (ppm), at different spectral resolutions of 1, 5, 10, and 20 wavelength bins (rows), for the three individual visits and an inverse-variance weighted average (columns). Each data point shows eclipse depth with 1$\sigma$ error bar on the y-axis and the width of the wavelength bin on the x-axis; depth uncertainties in the combined spectrum have been inflated to bring the three visits into consistency with each other. Each brown dashed line indicates the planet isothermal model as described in Equation \ref{eqn:single_temp} with uncertainty (brown shaded region) discussed in Equation \ref{eqn:model_unc}. The dark rock and thick atmosphere (grey) lines are Planck planet models calculated at 549\,K (0 albedo and instant reradiation; \( f = 2/3 \)) and 430\,K (0 albedo and perfect heat redistribution; \( f = 1/4 \)), respectively.}
\label{fig:emission_sparta}
\end{figure*}

With the eclipse timing and duration fixed to the values reported in Section \ref{subsec:ecosw_esinw} (translated back to $\sqrt{e}\cos\omega$ and $\sqrt{e}\sin\omega$), we obtained the emission spectra in terms of the eclipse depth ($D$), a planet-to-star flux ratio, from fits to broadband and spectroscopic light curves. Because eclipse depth is a nearly linear parameter and could go negative in these fits, the posteriors for eclipse depths were all well-described by symmetric Gaussian distributions. Figure \ref{fig:emission_sparta} and Table \ref{tab:depth_sparta} show these eclipse depths for the individual and combined visits at different binnings. Results are shown for \texttt{SPARTA} light curves, but the \texttt{Eureka!} depths are consistent to much better than 1 $\sigma$ for most bins (and no worse than 2$\sigma$ across any bins; see appendix \ref{appendix:depth_compare_sparta_eureka}). 

Larger wavelength bins have lower noise and average more strongly over possible interwavelength correlations, but they hide the details of the planetary emission that grows sharply across this wavelength range. Smaller wavelength bins are noisier, but they better reveal the shape of the emission spectrum and could potentially show atmospheric absorption features. Even though we observed broadly consistent depths across the wavelength range (5--12\,$\mu$m) as shown in Figure \ref{fig:emission_sparta} and Table \ref{tab:depth_sparta}, we decided to exclude data points at wavelengths longer than 10.6\,$\mu$m (grey shaded region in Figure \ref{fig:emission_sparta}) due to contamination from the shadowed region, and proceed with the more reliable shorter wavelength bins for comparison to theoretical models. This will make the comparison between each binning scheme more direct and easier to interpret. Therefore, the binning scheme is now 1, 4, 8, and 16 bins, instead of 1, 5, 10, and 20 bins as described in Section \ref{subsubsec:fixe}.

Also, instead of using individual visits in our statistical analysis and interpretation that might be too noisy, we combined the depths of all three visits using inverse-variance weighting. The skewed central wavelength of each data point in Figure \ref{fig:emission_sparta} indicates the ``effective wavelength'', the inverse-variance weighted-averaged wavelength. The effective wavelengths differ from the central bin wavelength the most in broadband and are less pronounced as the bin size decreases.

Overall, the emission spectra show good agreement between visits 1 and 3, with broadband eclipse depths of 37 $\pm$ 6\,ppm and 30 $\pm$ 8\,ppm, respectively, while visit 2 shows a deeper broadband eclipse depth at 62 $\pm$ 8\,ppm. We deployed the $\chi^2$ statistic to these three visits' broadband depths compared to the combined depth and got $\chi^2$ = 9.19 with 3-1 = 2 degrees of freedom (dof) which translated to $p$-value of 0.01. We assumed that all three measured eclipse depths were drawn from the same underlying distribution and artificially inflated the uncertainties of visits-combined broadband data point by a factor of $n_{\rm \sigma,combined}$ = $\sqrt{9.19/2} = 2.15$. This uncertainty inflation factor ($n_{\rm \sigma,combined}$) brings the reduced chi-square ($\chi^2_{\nu}$) of the three visits' broadband eclipse depths compared to the combined depth close to 1. Similarly, we calculated the uncertainty inflation ratio as described above for each binning scheme, which yielded factors of 1.72, 1.40, and 1.30 for the 4, 8, and 16 wavelength binning schemes, respectively. These inflation ratios met our expectation that smaller wavelength bin sizes exhibit more photon noise and are therefore less sensitive to systematic noise, resulting in smaller required $n_{\rm \sigma,combined}$. We then applied these uncertainty inflation ratios to the last column (combined) depths in Figure \ref{fig:emission_sparta} and Table \ref{tab:depth_sparta}.

For context, Figure \ref{fig:emission_sparta} shows the expected contrast for uniform, isothermal, Planck thermal emission from the planet's dayside as dashed gray lines, according to 
\begin{eqnarray}
    D_{\rm model} = \frac{\pi B_p(\lambda, T_p)}{F_{\ast}(T_{\rm \ast, eff}, \log g, [Fe/H])}\left(\frac{R_p}{R_{\ast}}\right)^2 \label{eqn:single_temp}
\end{eqnarray}
where $B_p(\lambda, T_p)$ is the Planck thermal emission intensity as a function of planet temperature ($T_p$) and wavelength ($\lambda$). $F_{\ast}(T_{\rm \ast, eff}, \log g, [Fe/H])$ indicates the SPHINX stellar spectrum as described in Section \ref{subsec:stellar_spec}. We also compared the approximation of a single dayside temperature to a sum of different temperature Planck spectra (from hottest at substellar point to coolest at the limb). We found that the difference is minimal: the sum of different temperature spectra is $\sim$1\% higher at 5\,$\mu$m and $\sim$3\% lower at 12\,$\mu$m. Therefore, we proceed with the single temperature approximation. 

In the calculation of the Planck spectra above, we did not yet consider the uncertainties associated with the system's parameters ($T_{\rm \ast, eff}$ = 3340 $\pm$ 150\,K, $a/R_{\ast}$ = 30.2 $\pm$ 1.7, $R_p/R_{\ast}$ = 0.0454 $\pm$ 0.0012; Table \ref{tab:parameters}). We translate these parameters into a model uncertainty ($\sigma_{model}$) and include them in the $\chi^2$ calculations as
\begin{eqnarray}
    \chi^2 = \frac{(D-D_{\rm model})^2}{\sigma_{\rm depth,combined}^2 + \sigma_{\rm model}^2},
    \label{eqn:chi2_model_unc}
\end{eqnarray}
where $D$ and $D_{\rm model}$ are the observed and modeled eclipse depths. $\sigma_{\rm depth,combined}$ is the inflated, combined-visits uncertainties on eclipse depths (as shown in the last column of Table \ref{tab:depth_sparta}) while $\sigma_{\rm model}$ represents the model uncertainties estimated via propagation of errors as
\begin{eqnarray}
    \sigma^2_{\rm model} &=& \left(\frac{\partial D_{\rm model}}{\partial T_{\ast}}\right)^2 \sigma^2_{T_{\ast}} + \left(\frac{\partial D_{\rm model}}{\partial \left(a/R_{\ast}\right)}\right)^2 \sigma^2_{\left(a/R_{\ast}\right)} \nonumber \\ 
    &+& \left(\frac{\partial D_{\rm model}}{\partial \left(R_p/R_{\ast}\right)}\right)^2 \sigma^2_{\left(R_p/R_{\ast}\right)},  \label{eqn:model_unc}
\end{eqnarray}
where the partial derivatives of the depth with respect to each parameter were evaluated numerically. Figure \ref{fig:emission_sparta} shows these model uncertainties as shaded swaths surrounding each model. Comparing to some key dayside temperature models for each binning, we find: 
\begin{itemize}
    \item The null hypothesis of there being no eclipse (0\,K Planck spectrum) can be ruled out at 4.6$\sigma$, 6.0$\sigma$, 7.7$\sigma$, and 9.2$\sigma$, for 1, 4, 8, and 16 bins. Therefore, we solidly detect the planet's thermal emission.
    \item The most basic thick-atmosphere hypothesis of zero-albedo, full heat-redistribution, and $f$ = 1/4 (431\,K Planck spectrum) can be marginally disfavored at 2.0$\sigma$, 2.3$\sigma$, 3.3$\sigma$, and 4.6$\sigma$ for 1, 4, 8, and 16 bins.
    \item The most basic no-atmosphere hypothesis of zero-albedo, no heat-redistribution, $f$ = 2/3 (549\,K Planck spectrum) is consistent with the data at 0.4$\sigma$, 0.3$\sigma$, 0.9$\sigma$, and 1.9$\sigma$ for 1, 4, 8, and 16 bins. 
\end{itemize} 

\begin{deluxetable*}{ccccccc}
\tablecaption{Eclipse depths for \texttt{SPARTA} pipeline}
\tablenum{2}
\label{tab:depth_sparta}
\tablehead{\colhead{$\lambda_{\text{eff}}$} & \colhead{$\lambda_{\text{edges}}$} & \colhead{$\lambda_{\text{c}}$} & \multicolumn{4}{c}{Eclipse depth (ppm)}\\ 
\colhead{($\mu$m)} & \colhead{($\mu$m)} & \colhead{($\mu$m)} & \colhead{visit 1} & \colhead{visit 2} & \colhead{visit 3} & \colhead{combined visits}  } 
%% All data must appear between the \startdata and \enddata commands
\startdata
Broadband &&&\\
7.1122 & [5.0759, 10.6228] & 7.8494 & 37 $\pm$ 6 & 62 $\pm$ 8 & 30 $\pm$ 8 & 41 $\pm$ 9 \\
\hline
5 wavelength bins &&&\\
5.7437 & [5.0759, 6.4607] & 5.7683 & 26 $\pm$ 10 & 58 $\pm$ 12 & 28 $\pm$ 12 & 36 $\pm$ 11 \\
7.1069 & [6.4607, 7.8455] & 7.1531 & 21 $\pm$ 12 & 57 $\pm$ 14 & 37 $\pm$ 15 & 36 $\pm$ 13 \\
8.4949 & [7.8455, 9.2304] & 8.5380 & 70 $\pm$ 13 & 49 $\pm$ 17 & 35 $\pm$ 16 & 54 $\pm$ 15 \\
9.8343 & [9.2304, 10.6152] & 9.9228 & 90 $\pm$ 18 & 114 $\pm$ 21 & 16 $\pm$ 21 & 76 $\pm$ 20 \\
11.1499 & [10.6152, 12.0000] & 11.3076 & 68 $\pm$ 39 & 130 $\pm$ 47 & 86 $\pm$ 56 & 91 $\pm$ 46 \\
\hline
10 wavelength bins &&&\\
5.4137 & [5.0759, 5.7683] & 5.4221 & 40 $\pm$ 14 & 66 $\pm$ 16 & 42 $\pm$ 17 & 49 $\pm$ 13 \\
6.1080 & [5.7683, 6.4607] & 6.1145 & 19 $\pm$ 14 & 54 $\pm$ 17 & 18 $\pm$ 16 & 29 $\pm$ 13 \\
6.7968 & [6.4607, 7.1531] & 6.8069 & 3 $\pm$ 15 & 70 $\pm$ 18 & 26 $\pm$ 21 & 30 $\pm$ 14 \\
7.4829 & [7.1531, 7.8455] & 7.4993 & 48 $\pm$ 16 & 44 $\pm$ 19 & 54 $\pm$ 20 & 49 $\pm$ 15 \\
8.1810 & [7.8455, 8.5379] & 8.1917 & 75 $\pm$ 17 & 64 $\pm$ 21 & 66 $\pm$ 22 & 70 $\pm$ 16 \\
8.8722 & [8.5379, 9.2304] & 8.8842 & 70 $\pm$ 20 & 42 $\pm$ 24 & -4 $\pm$ 23 & 39 $\pm$ 18 \\
9.5619 & [9.2304, 9.9228] & 9.5766 & 86 $\pm$ 21 & 116 $\pm$ 27 & 28 $\pm$ 27 & 78 $\pm$ 20 \\
10.2351 & [9.9228, 10.6152] & 10.2690 & 99 $\pm$ 27 & 106 $\pm$ 33 & 1 $\pm$ 34 & 74 $\pm$ 25 \\
10.9196 & [10.6152, 11.3076] & 10.9614 & 17 $\pm$ 43 & 206 $\pm$ 51 & 16 $\pm$ 55 & 73 $\pm$ 40 \\
11.6070 & [11.3076, 12.0000] & 11.6538 & 154 $\pm$ 67 & -44 $\pm$ 84 & 225 $\pm$ 101 & 108 $\pm$ 65 \\
\hline
20 wavelength bins &&&\\
5.2477 & [5.0759, 5.4221] & 5.2490 & 15 $\pm$ 19 & 80 $\pm$ 23 & 22 $\pm$ 24 & 36 $\pm$ 16 \\
5.5914 & [5.4221, 5.7683] & 5.5952 & 79 $\pm$ 19 & 54 $\pm$ 24 & 63 $\pm$ 24 & 68 $\pm$ 17 \\
5.9393 & [5.7683, 6.1145] & 5.9414 & 67 $\pm$ 19 & 66 $\pm$ 22 & 31 $\pm$ 24 & 57 $\pm$ 16 \\
6.2849 & [6.1145, 6.4607] & 6.2876 & -21 $\pm$ 19 & 50 $\pm$ 23 & 1 $\pm$ 23 & 6 $\pm$ 16 \\
6.6305 & [6.4607, 6.8069] & 6.6338 & -14 $\pm$ 22 & 79 $\pm$ 24 & 46 $\pm$ 35 & 32 $\pm$ 19 \\
6.9779 & [6.8069, 7.1531] & 6.9800 & 33 $\pm$ 20 & 61 $\pm$ 25 & 6 $\pm$ 34 & 37 $\pm$ 18 \\
7.3222 & [7.1531, 7.4993] & 7.3262 & 60 $\pm$ 21 & 36 $\pm$ 27 & 44 $\pm$ 29 & 49 $\pm$ 19 \\
7.6673 & [7.4993, 7.8455] & 7.6724 & 40 $\pm$ 23 & 49 $\pm$ 26 & 53 $\pm$ 28 & 46 $\pm$ 19 \\
8.0154 & [7.8455, 8.1917] & 8.0186 & 88 $\pm$ 23 & 50 $\pm$ 29 & 41 $\pm$ 31 & 65 $\pm$ 20 \\
8.3623 & [8.1917, 8.5379] & 8.3648 & 74 $\pm$ 23 & 92 $\pm$ 32 & 96 $\pm$ 33 & 84 $\pm$ 21 \\
8.7085 & [8.5379, 8.8842] & 8.7111 & 71 $\pm$ 27 & -24 $\pm$ 32 & -15 $\pm$ 35 & 20 $\pm$ 23 \\
9.0542 & [8.8842, 9.2304] & 9.0573 & 80 $\pm$ 28 & 126 $\pm$ 34 & 9 $\pm$ 34 & 73 $\pm$ 24 \\
9.4007 & [9.2304, 9.5766] & 9.4035 & 42 $\pm$ 27 & 100 $\pm$ 36 & 18 $\pm$ 36 & 51 $\pm$ 24 \\
9.7445 & [9.5766, 9.9228] & 9.7497 & 140 $\pm$ 31 & 142 $\pm$ 37 & 36 $\pm$ 39 & 113 $\pm$ 26 \\
10.0880 & [9.9228, 10.2690] & 10.0959 & 147 $\pm$ 33 & 140 $\pm$ 42 & 32 $\pm$ 44 & 116 $\pm$ 29 \\
10.4307 & [10.2690, 10.6152] & 10.4421 & 40 $\pm$ 39 & 58 $\pm$ 48 & -34 $\pm$ 52 & 27 $\pm$ 34 \\
10.7773 & [10.6152, 10.9614] & 10.7883 & 24 $\pm$ 51 & 138 $\pm$ 57 & 100 $\pm$ 60 & 82 $\pm$ 42 \\
11.1246 & [10.9614, 11.3076] & 11.1345 & 11 $\pm$ 68 & 334 $\pm$ 81 & -80 $\pm$ 103 & 97 $\pm$ 60 \\
11.4722 & [11.3076, 11.6538] & 11.4807 & 213 $\pm$ 78 & -60 $\pm$ 98 & 242 $\pm$ 119 & 135 $\pm$ 71 \\
11.8128 & [11.6538, 12.0000] & 11.8269 & 151 $\pm$ 100 & 63 $\pm$ 114 & 194 $\pm$ 142 & 130 $\pm$ 86 \\
\enddata
\tablecomments{Mean and standard deviation. Note that in last column (combined visits depths), the uncertainties got inflated so that $\chi^2_{\nu}$ of each visits depths compared to combined depths are 1.}
\tablecomments{Wavelength bins longer than 10.6 $\mu$m are included here for completeness but were excluded from atmospheric inferences.}
\end{deluxetable*}
  
\section{Discussion} \label{sec:discussions} % 3-4
Here we explore the implications of the detected eclipse and the planet's emission spectrum. We discuss LTT 1445A b's low eccentricity in context of the complex orbital dynamics of the multiplanet and triple-star system (Section \ref{subsec:eccentricity}), what the emission spectrum implies for possible atmospheric scenarios for the planet, both from the planet's overall energy budget perspective (Section \ref{subsec:dayside_temp}) and through comparison to atmospheric forward models (Section \ref{subsec:forward_model}), and we compare our findings with the ability of other planets within and beyond the Solar System to retain atmospheres (Section \ref{subsec:cosmic_shoreline}).

\subsection{Eccentricity in Context} \label{subsec:eccentricity}
The eccentricity of LTT 1445A b  is very close to zero. The $e\cos\omega$ and $e\sin\omega$ inferred from the eclipse timing and duration are each consistent at less than $2\sigma$ with 0, and we place an upper limit of $e < 0.0059$ at 95\% confidence from the visit-combined posterior distribution. 

In an ensemble analysis, \citet{sagear_orbital_2023} found that eccentricities of single transiting planets around M dwarfs can be large or small, whereas systems with multiple transiting planets around M dwarfs almost exclusively have small eccentricities. LTT 1445A b verifies this trend, now with an individually precisely measured small eccentricity in a compact multiple. \citet{quarles_orbital_2020} examined the planetary system's orbital stability in the face of perturbations from the BC component of the LTT 1445ABC triple and found the planets orbit tightly enough to be relatively unaffected by the two distant stars. The circular and coplanar orbits of the LTT 1445A planets point toward a cool dynamical history, free from major disruption by external perturbers, with any initial eccentricities for the planets likely erased by tidal circularization.

\subsection{Planet Dayside Temperature} \label{subsec:dayside_temp}
One way to assess the presence of the atmosphere on LTT 1445A b is through the overall energy budget: incoming radiation gets thermally reradiated from the planet's dayside, which is detectable at secondary eclipse. Here, we assume the planet's dayside is isothermal and infer a posterior distribution for its emission temperature. We define model depths ($D_{\rm model,binned}$) similar to Equation \ref{eqn:single_temp} but binned onto the same wavelength grid as the data by integrating the photon counts received by the planet and star within each bin:
\begin{eqnarray}
    D_{\rm model,binned} = \frac{\int \pi B_p E_{\lambda}^{-1} W_\lambda d\lambda }{\int F_{\ast} E_{\lambda}^{-1} W_\lambda d\lambda}\left(\frac{R_p}{R_{\ast}}\right)^2 \label{eqn:best_fit_temp}
\end{eqnarray}
where $E_{\lambda} = hc/\lambda$ is the energy of a photon at wavelength $\lambda$, converting fluxes from energy units to photon counts, which is appropriate for MIRI's photon-counting detector, and $W(\lambda)$ is the MIRI/LRS instrumental throughput\footnote{\url{https://jwst-docs.stsci.edu/jwst-exposure-time-calculator-overview/jwst-etc-pandeia-engine-tutorial}}, weighting wavelengths according to the efficiency with which JWST can count their photons. Uncertainties were propagated into the binned model depths as in Equation \ref{eqn:model_unc}. This integration is particularly important for wide wavelength bins across which the stellar flux drops dramatically and the planet's thermal emission rises sharply. We use \texttt{PyMC3} as described above to infer the dayside temperature $T_{\rm day}$ as one free parameter. 

\begin{figure}
\plotone{ 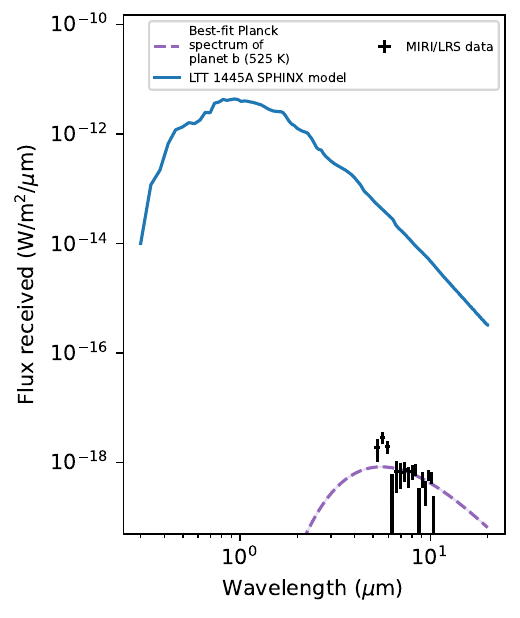}
\caption{The flux contrast between LTT 1445A's SPHINX model (solid blue line) and planet b's Planck thermal emission model at the best-fit temperature (525 K). The y-axis expresses the flux received at the telescope, given LTT 1445A's distance.}
\label{fig:flux_recieved}
\end{figure}

The best-fit dayside temperatures are $T_{\rm day}$ = 513 $\pm$ 33\,K, 511 $\pm$ 22\,K, 515 $\pm$ 17\,K, and 525 $\pm$ 15\,K for the 1, 4, 8, and 16 binning schemes, respectively. Figure \ref{fig:flux_recieved} shows the 16-bin spectrum converted into flux units along with the best-fit isothermal model, along with the stellar spectrum. The dayside temperature uncertainties do not account for uncertainties in the stellar and planetary parameters. To more accurately include them, we calculate a temperature ratio ($R$) between the fitted dayside brightness temperature ($T_{\rm day}$) and the maximum expected dayside temperature (assuming zero albedo and instant heat reradiation; $f_{\rm max}$ = 2/3) as 
\begin{eqnarray}
    R &=& \frac{T_{\rm day}}{T_{\rm max}} = \frac{T_{\rm day}}{(f_{\rm max}/f_{\rm eq})^{1/4}T_{eq}} \nonumber \\
    R &=& \frac{T_{\rm day}}{(8/3)^{1/4}T_{eq}} \label{eqn:R_ratio}
\end{eqnarray}
and propagated the uncertainty on $T_{\rm eq}$ = 431 $\pm$ 23\,K into the uncertainty on $R$, where $f_{\rm eq}$ = 1/4 is $f$-factor for zero albedo, fully redistribution. For 1, 4, 8, and 16 wavelength bins, we achieve a temperature ratio $R$ of 0.930 $\pm$ 0.077, 0.928 $\pm$ 0.064, 0.935 $\pm$ 0.059, and 0.952 $\pm$ 0.057.  Qualitatively, the $R$ uncertainties decreasing from 1 to 16 wavelength bins is consistent with the fact that a smaller uncertainty inflation ratio ($n_{\rm \sigma,combined}$) was applied to the higher-resolution combined emission spectra in Section \ref{subsec:emission_spectra}. The fitted $R$ ratios all agreed with each other within 1$\sigma$ level, and they are all consistent with instant reradiation from a dark rocky surface. Compared to the \texttt{SPARTA}-derived 16-bin temperature ratios of $R$ = 0.952 $\pm$ 0.057, for \texttt{Eureka!} we derive a comparable temperature ratio of $R$ = 0.944 $\pm$ 0.058, differing by only 0.8\% and well within the 1$\sigma$ uncertainty.

We then compared these temperature ratios ($R$) to simplified heat redistribution models from \citet{koll_scaling_2022}. An analytic model connects atmospheric surface pressure to the heat redistribution parameter ($f$-factor) with simplifying assumptions \citep{koll_scaling_2022}:

\begin{eqnarray}
    f = \frac{2}{3} - \frac{5}{12} \times \frac{\tau_{LW}^{1/3}\left(\frac{p_{\rm s}}{1 \text{bar}}\right)^{2/3}\left(\frac{T_{eq}}{600 \rm K}\right)^{-4/3}}{k + \tau_{LW}^{1/3}\left(\frac{p_{\rm s}}{1 \rm bar}\right)^{2/3}\left(\frac{T_{eq}}{600 \rm K}\right)^{-4/3}} \label{eqn:f_factor}
\end{eqnarray}
where $p_{\rm s}$ represents the surface pressure. We linearly interpolated equivalent long-wave gray optical thickness $\tau_{LW}$ as a function of $p_{\rm s}$ from $\tau_{LW}$ = [0.2170, 0.5533, 1.1497, 3.5244, 7.0824] calculated assuming pure CO$_2$ atmosphere in Section \ref{subsec:forward_model} when $p_{\rm s}$ = [0.01, 0.1, 1.0, 10.0, 100.0] bar, respectively.

\[ k = \frac{Lg}{\chi \beta c_p} \times \left(\frac{C_d \sigma^2}{R}\right)^{1/3}(1 \rm bar)^{-2/3} (600 \rm K)^{4/3} \label{eqn:k_koll}\]
captures additional planetary properties that have minimal effects across different planetary systems (mostly of order unity). We adopt parameter values from \citet{koll_scaling_2022} for high-MMW atmospheres: ($R$, $c_p$) = ($R_{N_2}$, $c_{p_{N_2}}$), $C_d = 1.9 \times 10^{-3}$, $L = R_p$, $\chi = 0.2$, $\beta \equiv \frac{c_p}{R n_{LW}}$ where $n_{LW} = 2$ indicates opacity is dependent on pressure (pressure broadening). Additionally, we calculated $g = 14.93\,m\,s^{-2}$ for LTT 1445A b (using parameters from Table \ref{tab:parameters}), yielding $k_{\rm LTT 1445A b}$ = 2.77. We used Equations \ref{eqn:f_factor} and \ref{eqn:dayside_temp_with_albedo} to calculate the expected dayside temperature, although the model predictions are robust to uncertainties in the system parameter in Table \ref{tab:parameters}. The exact quantitative relationship between surface pressure and dayside temperature depends on the details of the above parameter assumptions. 

\begin{figure}
\plotone{ 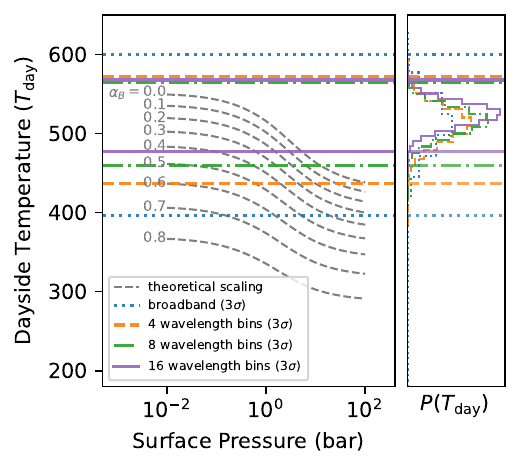}
\caption{Analytical dayside temperature estimates as a function of surface pressure and Bond albedo ($\alpha_{\rm B}$) from \citet{koll_scaling_2022} (grey dashed lines in left panel) with observed dayside temperature posteriors from different binning strategies (right panel). Horizontal lines indicate 3$\sigma$ temperature limits from the observations. The uncertainties shown here do not account for uncertainties in the system's parameters.}
\label{fig:koll_albedo_hist}
\end{figure}

Figure \ref{fig:koll_albedo_hist} shows these calculated temperatures as a function surface pressure ($p_{\rm s}$), for different assumed Bond albedos ($\alpha_{\rm B}$), compared to the planet's dayside temperature posteriors inferred from this work. In this model, the planet's dayside might be cooler either because an atmosphere distributes the incoming radiation to the nightside or because the planet surface ``or'' reflective cloud-deck reflects some of the incoming light before it can be absorbed. For context, \citet{mansfield_identifying_2019} identified $\alpha_{\rm B} \approx 0.4$ as an approximate upper limit for the expected albedo of warm rocky surfaces; in Figure \ref{fig:koll_albedo_hist} that implies that a dayside temperature cooler than 483\,K could only be achieved through an atmosphere. At all binnings, the observed dayside temperatures are warmer than this limit. The dayside temperature disfavors atmospheres that are extremely reflective and/or very thick, but many atmospheric scenarios cannot be explicitly distinguished from bare rock, such as moderately thick (Earth-like) and thin (Mars-like) atmospheres.

\begin{figure}
\plotone{ 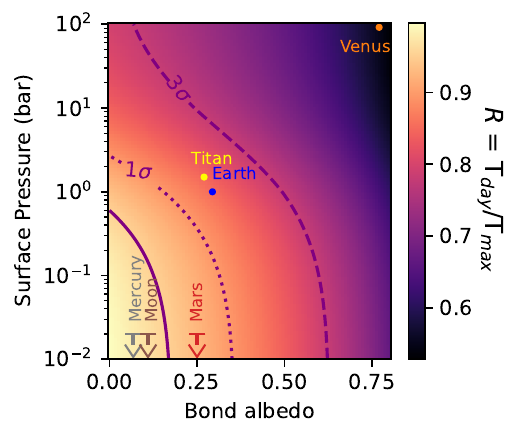}
\caption{A map of expected dayside temperature ratio ($R = T_{\rm day}/T_{\rm max}$) as a function of Bond albedo and surface pressure, according to the \citet{koll_scaling_2022} analytic estimates. Contours indicate LTT 1445A b's temperature ratio of $R$ = 0.952 $\pm$ 0.057 measured from the 16-bin emission spectrum, showing the central value (solid line), $1\sigma$ range (dotted line), and $3\sigma$ range (dashed line), accounting for both measurement uncertainties in the emission spectrum and underlying uncertainties in the system parameters. Several Solar System planetary bodies are shown for reference.}
\label{fig:koll_albedo_heat_map}
\end{figure}

Figure \ref{fig:koll_albedo_heat_map} shows the analytic temperature calculation expressed as the temperature ratio $R$, as a 2D map of surface pressure and albedo. Posteriors for the observed temperature ratio $R$ = 0.952 $\pm$ 0.057 (from the 16 bin fit) are shown, where the uncertainty includes both JWST measurement uncertainties and from the intrinsic system parameter uncertainties. The observed temperature ratio disfavors thick ($p_{\rm s} >$ 100 bar) atmospheres with Bond albedos $>$ 0.08 at the 3$\sigma$ confidence interval. These thick atmospheres too effectively transport heat to the night side, thus cooling the dayside below the observed temperature ratio. Additionally, the temperature ratio rules out a bright ($\alpha_{\rm B} > 0.62$) bare-rock at 3$\sigma$ confidence with the same logic. 

We include Solar System rocky worlds in Figure \ref{fig:koll_albedo_heat_map} for comparison. Even though their Bond albedos would be different if irradiated by a cooler M dwarf compared to the Sun, a qualitative comparison is still beneficial for understanding the atmospheric possibilities on LTT 1445A b. Venus has a thick ($p_{\rm s}$ = 92 bar)\footnote{\label{footnote:fact_sheet} \url{https://nssdc.gsfc.nasa.gov/planetary/planetfact.html}} mainly CO$_2$ atmosphere with reflective clouds ($\alpha_{\rm B}$ = 0.77)$^{\ref{footnote:fact_sheet}}$. A true Venus-like atmosphere is not aligned with our derived surface pressure and Bond albedo from the observations and can be ruled out at $\sim$7 $\sigma$ confidence. Titan with $\alpha_{\rm B}$ = 0.265 \citep{li_global_2011}, $p_{\rm s}$ = 1.5 bar \citep{lindal_atmosphere_1983} and Earth ($\alpha_{\rm B}$ = 0.294, $p_{\rm s}$ = 1 bar)$^{\ref{footnote:fact_sheet}}$ are good examples of moderate surface pressures and albedos; we slightly disfavor such atmospheres at $\sim$ 1.6$\sigma$ but cannot rule them out. Thin atmospheres like the one on Mars ($\alpha_{\rm B}$ = 0.250, $p_{\rm s} = [4.0-8.7] \times 10^{-3}$ bar)$^{\ref{footnote:fact_sheet}}$ or non-existent atmospheres like that of Mercury ($\alpha_{\rm B}$ = 0.068, $p_{\rm s} \lesssim 5 \times 10^{-15}$ bar)$^{\ref{footnote:fact_sheet}}$ and the Moon ($\alpha_{\rm B}$ = 0.11, $p_{\rm s} \lesssim 3 \times 10^{-15}$ bar)$^{\ref{footnote:fact_sheet}}$ are all consistent with the observed temperature ratio within 1$\sigma$. 

It is worth noting that we observed statistically different emission spectra from each visit, especially the second visit (shown in Figure \ref{fig:emission_sparta}). Such differences might result either from uncorrected systematics or from astrophysical time variability on the planet's flux. We accounted for the inconsistent depths by inflating the emission spectrum uncertainties, but if we had used only the two most consistent visits (1 + 3), the resulting lower eclipse depth would imply a cooler temperature ratio of $R=0.914 \pm 0.057$. This would still reject reflective Venus-like atmospheres at $\sim$ 6$\sigma$ but it would not reject thick ($\sim$ 100 bar) but dark ($\alpha_{\rm B} < 0.25$) atmosphere from an overall energy budget perspective.

  \subsection{Forward model atmospheres} \label{subsec:forward_model}
In addition to the overall energy balance argument above, an atmosphere might be revealed through the detection of molecular absorption features in the emission spectrum. Here, we compare the MIRI/LRS emission spectrum of LTT 1445A b to forward models of simple atmospheres. We constructed model emission spectra using calculations as in \citet{morley_observing_2017}, assuming 100\% CO$_2$ atmospheres of varying surface pressures of [0.01, 0.1, 1.0, 10.0, 100.0] bar. Temperature-pressure profiles follow a dry adiabat from the surface up until they reach the skin temperature, above which they are isothermal. or all models, the Bond albedo was assumed to be $\alpha_B = 0.1$ for both the overall energy balance of the atmosphere and for the planet’s surface. The \citet{koll_scaling_2022} analytic heat redistribution is included with a self-consistent infrared opacity ($f$ = [0.662, 0.619, 0.434, 0.272, 0.252] for $p_{\rm s}$ = [0.01, 0.1, 1.0, 10.0, 100.0] bar, respectively), so thicker atmospheres have lower $f$, distributing more energy to the nightside and thus requiring less total heat be emitted from the dayside visible during eclipse.

\begin{figure*}
\plotone{ 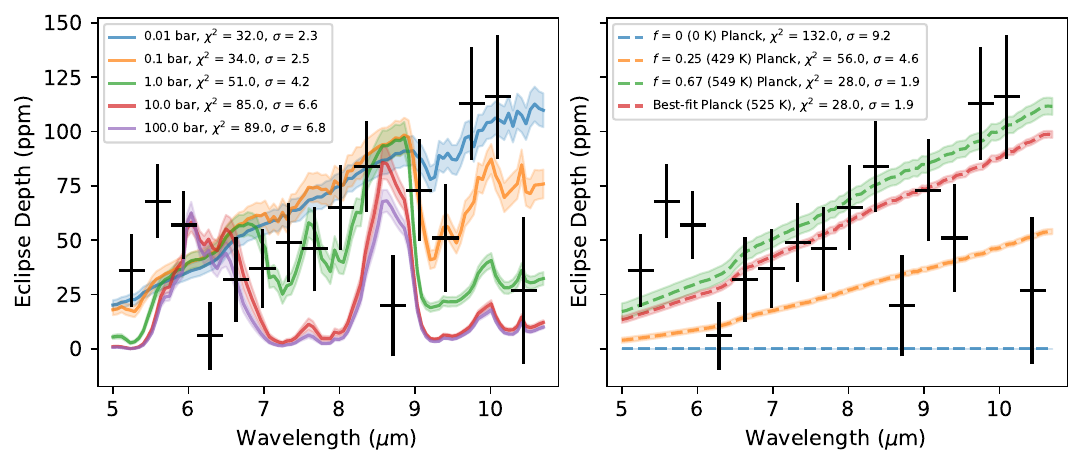}
\caption{The average emission spectrum of LTT 1445A b compared to models with and without atmospheres. (Left) Forward models are shown for 100\% CO$_2$, $\alpha_{\rm B}$ = 0.1 atmospheres with at different surface pressures ($p_{\rm s}$). (Right) Planck thermal spectra are shown at 0\,K (no eclipse), 431\,K ($\alpha_{\rm B}$ = 0, $f$-factor = 1/4), 549\,K ($\alpha_{\rm B}$ = 0, $f$-factor = 2/3), and best-fit dayside temperature (525 K).  The displayed $\chi^2$ incorporates model uncertainty from the uncertain system parameters, is for 16 degreess of freedom, and is translated into a probability with which each model can be disfavored, expressed as a number of $\sigma$ in a Gaussian distribution.}

\label{fig:forward_model_chi2}
\end{figure*}

Figure \ref{fig:forward_model_chi2} compares these model spectra to our 16-bin measured emission spectrum. For each model, we calculate the $\chi^2$ statistic including the model uncertainty using Equation~\ref{eqn:chi2_model_unc}, where the numerical derivatives needed for Equation~\ref{eqn:model_unc} were calculated from model emission spectra generated for two closely separated equilibrium temperatures. Comparing the achieved $\chi^2$ to the probability distribution for $\chi^2$ assuming 16 degrees of freedom, we calculate the probability at which a given model can be disfavored and translate that into a number of $\sigma$ away from a Gaussian distribution. 

Figure \ref{fig:forward_model_chi2} shows the thinner atmospheres are most consistent with the data, with emission spectra that are effectively closer to the warm isothermal spectrum explored in Section \ref{subsec:dayside_temp}. Surface pressures $\le 0.1$ bar do not perfectly match to the data, but are still allowed within $3\sigma$. For these simple atmospheres, thicker $\ge 1$ bar surface pressures are progressively worse fits, with stronger absorption features resulting in too little flux in CO$_2$ bands and too much flux in atmospheric windows where hotter temperatures from deep in the atmosphere shine through; the tested atmospheres are all disfavored at $>4 \sigma$.

Taken together, the lack of deep CO$_2$ absorption features (Figure \ref{fig:forward_model_chi2}) and the dayside temperature being broadly consistent with instant reradiation from a rocky surface (Figure \ref{fig:koll_albedo_heat_map}), point toward disfavoring the presence of a thick $\sim 100$ bar atmosphere on LTT 1445A b. An important caveat is that we only explored the most basic atmospheric forward models with pure CO$_2$ composition and monotonically decreasing temperature pressure profiles. Such simple modeling is reasonable given that the data are consistent with isothermal emission spectra at $2\sigma$ (Figure \ref{fig:forward_model_chi2}). However, it is possible that a more complete exploration of atmospheric compositions and thermal structures might better explain the marginally-significant low-level wiggles in the emission spectrum. For example, the shallow depths at 6.3 and 8.7 $\mu$m align with atmospheric CO$_2$ opacity windows; a moderate dayside temperature inversion could potentially better match the data, with high emission in wavelengths of high CO$_2$ opacity and cooler emission from the deeper atmosphere. Fundamentally, more precise data {\em and} more detailed atmospheric modeling will likely be required to definitely characterize whatever atmosphere might remain around LTT 1445A b.

\subsection{Placing LTT 1445A b in Context} \label{subsec:cosmic_shoreline}

\begin{figure}
\plotone{ 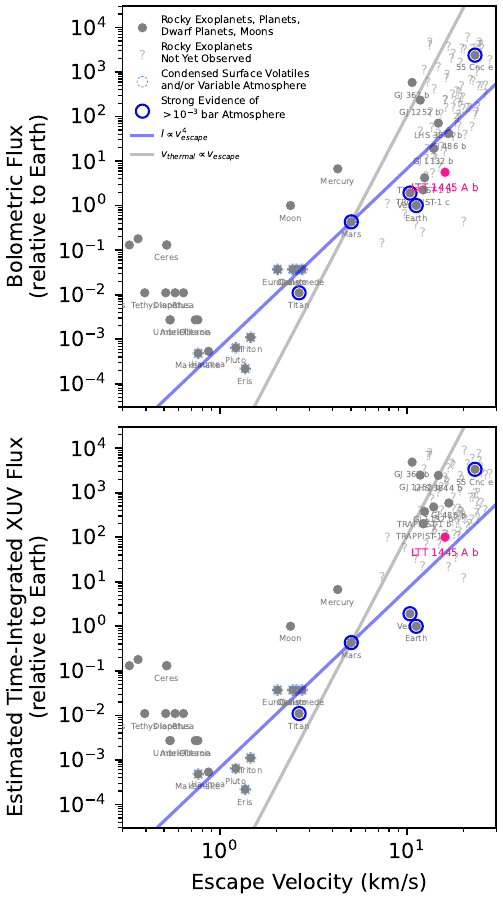}
\caption{LTT 1445A b in the context of other worlds with and without atmospheres, considering both instantaneous bolometric instellation (top) and a crudely estimated proxy for time-integrated high-energy received (bottom). Two relevant boundaries are shown, each normalized to Mars: an empirical $I\propto v_{escape}^4$  ``cosmic shoreline'' proposed by \citet{zahnle_cosmic_2017} and a simpler $v_{thermal}\propto v_{escape}$ energy-balance shoreline.}
\label{fig:cosmic_shoreline}
\end{figure}

We compare LTT 1445A b, a rocky planet for which MIRI/LRS disfavors thick Venus-like atmospheres, with other worlds with and without atmospheres. Figure \ref{fig:cosmic_shoreline} presents proposed ``cosmic shorelines,'' potentially separating planets with and without atmospheres \citep{zahnle_cosmic_2017}. Exoplanet data were drawn from the NASA Exoplanet Archive\footnote{\url{https://exoplanetarchive.ipac.caltech.edu/}} \texttt{pscomppars} table, and limited to systems with better than 20\% uncertainties on the planet mass and radius, curated to replace bad parameter choices for a few important nearby systems, and filtered to include only planets smaller than 2 R$_\oplus$ to exclude most thick H/He-dominated envelopes, focusing on secondary rocky atmospheres. Solar System data for planets, satellites, and small bodies were drawn from JPL Solar System Dynamics\footnote{\url{https://ssd.jpl.nasa.gov}}, again eliminating (the four) gas and ice giants with thick H/He-envelopes. 

Bolometric flux and escape velocity are calculated straightforwardly from system parameters. The ``estimated time-integrated XUV flux'' is calculated via the commonly-used but extremely over-simplified approximation from \citet{zahnle_cosmic_2017} that the cumulative XUV energy output of a star $\int L_{\rm XUV} dt$ is set by a proportionality with the star's instantaneous bolometric luminosity as $\propto L_{\rm bol}^{0.4}$. Qualitatively, this very rough tracer has the effect that lower-luminosity M dwarf planets appear additionally more highly irradiated. Quantitatively, this cumulative XUV tracer approximating polynomial scalings from \citet{lammer_determining_2009} has enormous uncertainties because it is derived from incomplete X-ray data extrapolated into the mostly unobserved EUV, it does not incorporate the intrinsic spread in XUV fluxes at fixed age, and it does not actually use individual exoplanet ages so that young and old planets would inaccurately appear to have absorbed the same integrated XUV over their lifetimes. Given these concerns, relative XUV histories might easily be uncertain to more than an order of magnitude, possibly including systematic trends and should be viewed with utmost skepticism.

Figure \ref{fig:cosmic_shoreline} draws dark blue circles around planets if there is strong evidence of the presence of an atmosphere thicker than 1 mbar. For the Solar System the only such rocky atmospheres are Venus, Earth, Mars, and Titan, and for the exoplanets this so far includes only 55 Cnc e \citep{hu_secondary_2024, patel_jwst_2024} which sits at extremes in both irradiation and size. Exoplanets that have been observed in thermal emission with JWST or Spitzer to test for atmospheres are marked as solid circles without blue circles (even if observations might not have been sensitive to all mbar atmospheres), while exoplanets with absolutely no observational atmosphere constraints are shown as question marks. Solar System bodies with atmospheres in vapor pressure equilibrium with large reservoirs of condensed surface volatiles like Pluto are denoted with dashed outlines, as are transient tenuous atmospheres fed by volcanism like the Galilean moons. 

Power laws in each panel of Figure \ref{fig:cosmic_shoreline} aim to delineate planets that have atmospheres from those without, each normalized to Mars as a planet that has experienced significant but incomplete atmospheric loss \citep{jakosky_loss_2018}. The blue lines are the proposed empirical $I\propto v_{\rm escape}^4$ cosmic shorelines from \citet{zahnle_cosmic_2017}, originally drawn to connect hot hydrogen-rich exoplanets with tenuous cold atmospheres like Pluto. The gray lines indicate another simple shoreline set by a constant ratio between the planet's escape velocity $v_{\rm escape}$ and the thermal velocity $v_{\rm thermal}$ of atmospheric molecules, effectively a line of constant Jeans escape parameter $\lambda = E_{\rm gravitational}/E_{\rm thermal}$ quantifying whether gas particles have the kinetic energy to escape the planet's gravity. With $v_{\rm thermal} \propto T^{1/2}$ and equilibrium temperatures set by insolation as $T \propto I^{1/4}$, constant values of $v_{\rm escape}/v_{\rm thermal}$ will follow $I\propto v_{\rm escape}^8$, thus appearing steeper in the plots. Thermal velocities calculated from the bolometric flux represents particle speeds for the planet's equilibrium temperature, which is reasonably well known (assuming similar albedos) but might be much cooler than the temperatures in the exosphere where escape occurs; using average XUV irradiation might be a better tracer of exospheric heating, but it relies on uncertain inputs and neglects the difficult-to-predict balance with exospheric cooling. 

Venus or Earth have roughly 100 bars of CO$_2$ (considering the CO$_2$ locked by liquid water into Earth's global limestone deposits). LTT 1445 A b's apparent lack of such a thick CO$_2$ atmosphere makes more sense in Figure \ref{fig:cosmic_shoreline}'s bottom panel, where its enhanced XUV moves it farther up, away from the Solar System planets. We observe that LTT 1445A b is below the Mars-normalized $I \propto v^4_{\rm escape}$ shoreline in the bolometric panel, but above it in the time-integrated XUV panel. This may potentially point toward the unsurprising conclusion that XUV irradiation matters for atmospheric evolution, but we caution against too strongly interpreting these simple shorelines since they neglect many important factors governing atmospheric retention and delivery (stellar winds, magnetic fields, chemistry, impacts, tides, outgassing).

\citet{diamond-lowe_high-energy_2024} investigated LTT 1445 A's current high-energy spectrum in the X-ray and UV using data from Chandra and Hubble COS + STIS. Though no flares have been seen in the optical, flares were detected at both UV and X-ray wavelengths. Given the observed XUV activity and their poorly constrained lower limit age of 2.2 Gyr, they proposed that LTT 1445A b would be able to retain a pure CO$_2$ atmosphere if it started with 10\% or more of Earth's total CO$_2$ budget. If LTT 1445A b started with a total CO$_2$ budget that was less than 10\% of Earth's total CO$_2$ budget (according to their simplified model), it would not be able to retain its CO$_2$ atmosphere, and that might be one of the reasons for the observed lacking Venus-like atmosphere.

\section{Conclusion} \label{sec:conclusion} % 0.5 pages
In this work, we obtained three JWST MIRI/LRS eclipses of LTT 1445A b to construct a thermal emission spectrum for this nearby rocky exoplanet. We detected the secondary eclipse with an eclipse time 2.4 $\pm$ 1.4 minutes later than expected for a circular orbit. Combining the three eclipses together, the broadband 5--10.6\,$\mu$m eclipse depth of $41 \pm 9$\,ppm and the same data split into emission spectra spanning those wavelengths with 4, 8, or 16 bins are all consistent with instant reradiation of incoming stellar energy from a hot planet dayside. This bright dayside emission is consistent with emission from a dark rocky surface, and it disfavors a thick, 100-bar, Venus-like CO$_2$ atmosphere. From an energy balance perspective, the average measured dayside brightness temperature is too hot to be easily explained by a substantial atmosphere able to circulate heat efficiently to the planet's nightside ($>10$ bar for Earth-like albedos; Figure \ref{fig:koll_albedo_heat_map}). From a comparison with model emission spectra, the lack of deep CO$_2$ absorption features is inconsistent with a strongly greenhouse-warmed surface emitting through atmospheric windows; the data exclude simple $>1$ bar CO$_2$ atmospheres at $>4\sigma$ (Figure \ref{fig:forward_model_chi2}), but with a caveat that we did not broadly explore all possible compositions or thermal structures. The apparent lack of a thick CO$_2$ atmosphere might be a result of atmospheric loss due to strong XUV upper atmosphere heating or any number of other loss processes that are difficult to quantify for exoplanet systems. 

These MIRI/LRS observations do not altogether rule an atmosphere for LTT 1445 A b. Moderate atmospheres of various compositions with up to about 10 bars of surface pressure are still allowed by the data but would require further analyses and/or observations to probe. We suggest the following future work to more definitively address possible atmospheres on LTT 1445A b:

\begin{itemize}
    \item For the detection of absorption features, additional thermal eclipse observations could resolve the current ambiguity about LTT 1445 A b's atmosphere, particularly if they are sensitive to atmospheric absorption features at complementary wavelengths. Eclipses with MIRI filter photometry at 12.8\,$\mu$m and 15\,$\mu$m would span a deep CO$_2$ absorption feature, potentially improving sensitivity to lower surface pressures than we can address here. 
    \item For the energy budget argument, the uncertainty on our calculated temperature ratio $R = T_{\rm day}/T_{\rm max}$ includes about equal contributions from the uncertainty on the measured MIRI/LRS dayside temperature $T_{\rm day}$ and from the uncertainties propagated from the system parameters into the predicted maximum dayside temperature $T_{\rm max}$. As JWST transits can vastly improve precision on both planetary and stellar parameters \citep{eastman_beating_2023,mahajan_using_2024}, a reanalysis incorporating the upcoming COMPASS transit of LTT 1445A b (JWST-GO-2512) or other precise transits could shrink uncertainties on the stellar $T_{\rm \ast, eff}$, $R_p/R_{\ast}$, $a/R_{\ast}$ and thus improve what statements we can make about LTT 1445A b's atmospheric recirculation.     
    \item For future data analysis, we needed to scale up the depth uncertainties to account for differences among the three eclipses that were larger than their individual uncertainties ($2.15\times$ for broadband, $1.30\times$ for 16 wavelengths). Revisiting these existing data with a more accurate model for MIRI/LRS instrumental systematics, especially the shadowed region, could therefore potentially shrink the depth uncertainties by 30\% relative to those shown here.
    \item For future modeling, the atmospheric models shown here assumed simple CO$_2$ compositions with monotonically decreasing temperature-pressure profiles, not exploring the effects of photochemistry or dayside temperature inversions. Figure \ref{fig:forward_model_chi2} shows tantalizing hints of cool planet emission aligning with CO$_2$ atmospheric windows that might be better explained with an inverted temperature profile, although the current MIRI/LRS emission spectrum's consistency with isothermal emission limits the detection of any molecular features to 2$\sigma$ at best. 
\end{itemize}
LTT 1445Ab is definitely not just like Venus, with have a thick atmosphere cooling its dayside emission through both reflective clouds and recirculation of heat away across the planet. However, with many atmospheric scenarios still possible, LTT 1445A b remains a compelling target for mapping the ability of rocky planets to retain atmospheres in the face of the harsh M dwarf stellar environment. 

\section{Acknowledgements}
The authors want to thank the JWST program coordinator Tricia Royle and MIRI reviewer Greg Sloan at Space Telescope Science Institute (STScI) for the planning of this observation. Support for program JWST-GO-2708 was provided by NASA through a grant from the Space Telescope Science Institute, which is operated by the Associations of Universities for Research in Astronomy, Incorporated, under NASA contract NAS5- 26555. terial is based upon work by PW, ZKBT, and CM supported by the National Science Foundation under Grant No. 1945633.  HDL acknowledges support from the Carlsberg Foundation, grant CF22-1254.This research has made use of the NASA Exoplanet Archive, which is operated by the California Institute of Technology, under contract with the National Aeronautics and Space Administration under the Exoplanet Exploration Program. This work utilized the Alpine high performance computing resource at the University of Colorado Boulder. Alpine is jointly funded by the University of Colorado Boulder, the University of Colorado Anschutz, and Colorado State University.

\facilities{JWST (MIRI/LRS), Exoplanet Archive}

\section{Data availability}
This work used the data from Mid-Infrared Instrument (MIRI) onboard James Webb Space Telescope (JWST) under cycle 1 GO (proposal 2708; PI: Zach Berta-Thompson).
\section{Code availability}
We used publicly available code \texttt{Eureka!}\footnote{\citet{bell_eureka_2022}}, \texttt{SPARTA}\footnote{\citet{zhang_gj_2024}}, \texttt{chromatic}\footnote{\url{https://github.com/zkbt/chromatic/}}, \texttt{chromatic\_fitting}\footnote{\url{https://github.com/catrionamurray/chromatic\_fitting}}, \texttt{exoplanet-atlas}\footnote{\url{https://github.com/zkbt/exoplanet-atlas}}, \texttt{starry}\footnote{\citet{luger_starry_2019}}, \texttt{PyMC3}\footnote{\citet{salvatier_probabilistic_2016}}.

\bibliography{LTT1445Ab}{}
\bibliographystyle{aasjournal}


\section{Appendix}

  \subsection{Planet flux comparison between \texttt{Eureka!} and \texttt{SPARTA} pipeline} \label{appendix:depth_compare_sparta_eureka}
  While not showing in the main figures, both \texttt{SPARTA} and \texttt{Eureka!} data sets at 16 wavelength bins show planet emission spectra that 
all agree with 2$\sigma$, with most data points agreeing within 1$\sigma$ except 9 out off 80 points. If we further exclude points within shadowed region, only 3 points that is not agreed within 1$\sigma$ but rather 2$\sigma$ and all of the points in combined visits agreed within 1$\sigma$. This result emphasized the robustness of the emission spectra.
\begin{figure*}[h]
        % Include an image, with a width that's half the width allowed for text.
        \centering
\plotone{ 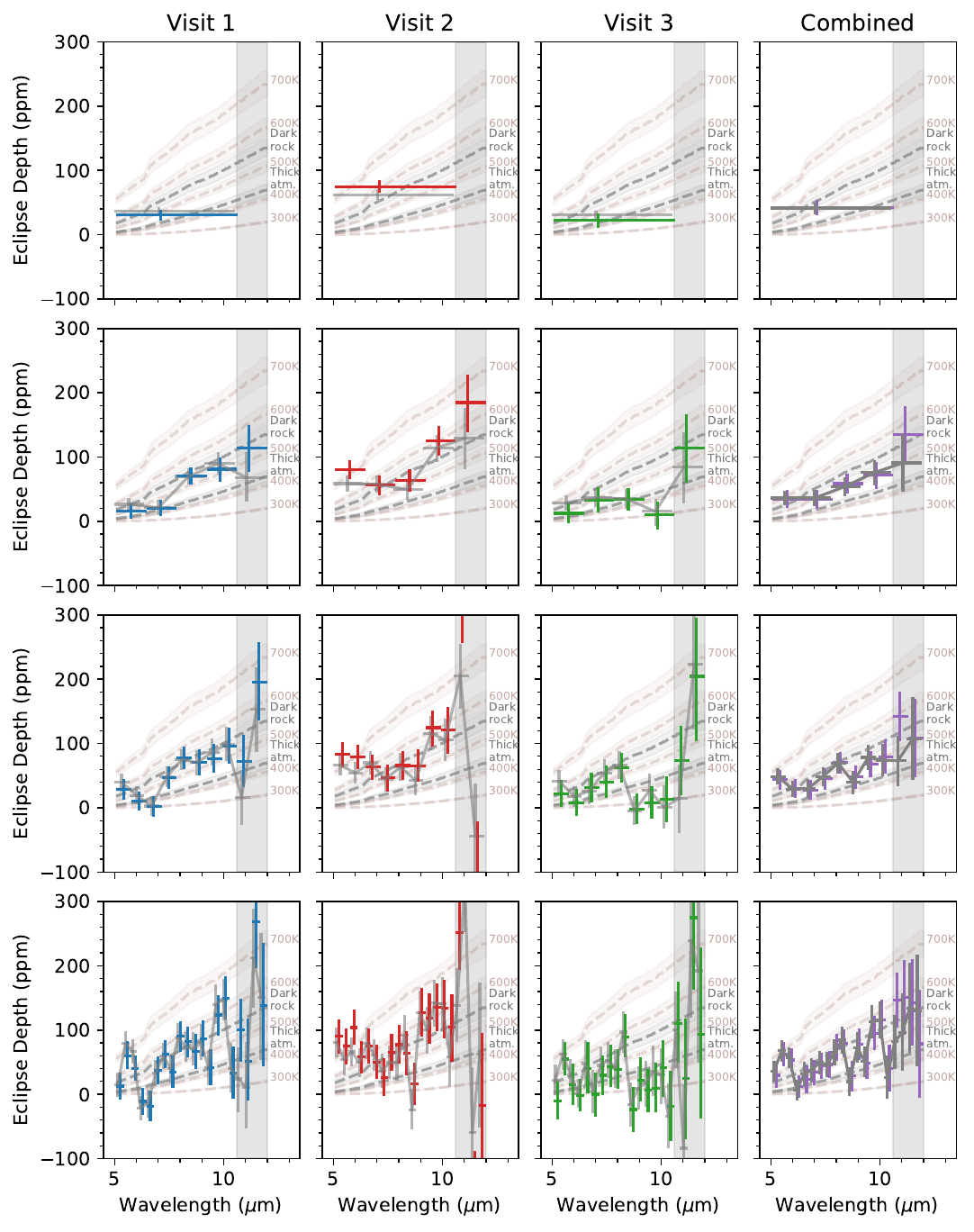}
\caption{Emission spectra comparison in terms of measured secondary eclipse depth (ppm). We constructed this plot in the same fashion as Figure \ref{fig:emission_sparta}. The grey errorbars indicate eclipse depths from \texttt{SPARTA} pipeline and were offset in by -0.1 micron in x-axis for better visualization while colored errorbars showed eclipse depths from \texttt{Eureka!} pipeline.}
        % Include a caption for this figure.
        
    \end{figure*}

\begin{deluxetable*}{ccccccc}
\tablecaption{Eclipse depth for \texttt{Eureka!} pipeline}
\tablenum{2}
\label{tab:depth_eureka}
\tablehead{\colhead{$\lambda_{\text{eff}}$} & \colhead{$\lambda_{\text{edges}}$} & \colhead{$\lambda_{\text{c}}$} & \multicolumn{4}{c}{Eclipse depth (ppm)}\\ 
\colhead{($\mu m$)} & \colhead{($\mu m$)} & \colhead{($\mu m$)} & \colhead{visit 1} & \colhead{visit 2} & \colhead{visit 3} & \colhead{combined visits}  } 

%% All data must appear between the \startdata and \enddata commands
\startdata
Broadband &&&\\
7.1158 & [5.0774, 10.6246] & 7.8510 & 31 $\pm$ 8 & 75 $\pm$ 10 & 22 $\pm$ 11 & 43 $\pm$ 16 \\
\hline
5 wavelength bins &&&\\
5.7460 & [5.0774, 6.4623] & 5.7698 & 16 $\pm$ 12 & 80 $\pm$ 15 & 12 $\pm$ 16 & 35 $\pm$ 17 \\
7.1117 & [6.4623, 7.8472] & 7.1547 & 20 $\pm$ 13 & 56 $\pm$ 15 & 32 $\pm$ 19 & 34 $\pm$ 18 \\
8.4941 & [7.8472, 9.2321] & 8.5397 & 71 $\pm$ 14 & 63 $\pm$ 17 & 34 $\pm$ 17 & 59 $\pm$ 19 \\
9.8325 & [9.2321, 10.6171] & 9.9246 & 81 $\pm$ 19 & 125 $\pm$ 23 & 10 $\pm$ 22 & 72 $\pm$ 25 \\
11.1422 & [10.6171, 12.0020] & 11.3095 & 114 $\pm$ 36 & 185 $\pm$ 46 & 115 $\pm$ 54 & 135 $\pm$ 51 \\
\hline
10 wavelength bins &&&\\
5.4176 & [5.0774, 5.7698] & 5.4236 & 29 $\pm$ 17 & 83 $\pm$ 20 & 22 $\pm$ 21 & 43 $\pm$ 18 \\
6.1114 & [5.7698, 6.4623] & 6.1161 & 11 $\pm$ 15 & 79 $\pm$ 19 & 8 $\pm$ 21 & 30 $\pm$ 17 \\
6.7976 & [6.4623, 7.1547] & 6.8085 & 2 $\pm$ 16 & 64 $\pm$ 20 & 32 $\pm$ 24 & 28 $\pm$ 18 \\
7.4852 & [7.1547, 7.8472] & 7.5010 & 47 $\pm$ 18 & 46 $\pm$ 21 & 39 $\pm$ 26 & 45 $\pm$ 20 \\
8.1819 & [7.8472, 8.5397] & 8.1934 & 78 $\pm$ 18 & 66 $\pm$ 22 & 62 $\pm$ 23 & 70 $\pm$ 19 \\
8.8726 & [8.5397, 9.2321] & 8.8859 & 71 $\pm$ 20 & 66 $\pm$ 25 & -2 $\pm$ 24 & 48 $\pm$ 21 \\
9.5632 & [9.2321, 9.9246] & 9.5784 & 77 $\pm$ 22 & 125 $\pm$ 28 & 8 $\pm$ 26 & 68 $\pm$ 23 \\
10.2351 & [9.9246, 10.6171] & 10.2708 & 95 $\pm$ 28 & 121 $\pm$ 37 & 14 $\pm$ 36 & 80 $\pm$ 31 \\
10.9164 & [10.6171, 11.3095] & 10.9633 & 73 $\pm$ 41 & 308 $\pm$ 51 & 74 $\pm$ 54 & 142 $\pm$ 45 \\
11.6060 & [11.3095, 12.0020] & 11.6558 & 195 $\pm$ 62 & -102 $\pm$ 80 & 206 $\pm$ 98 & 109 $\pm$ 71 \\
\hline
20 wavelength bins &&&\\
5.2495 & [5.0774, 5.4236] & 5.2505 & 13 $\pm$ 22 & 91 $\pm$ 27 & -11 $\pm$ 30 & 30 $\pm$ 20 \\
5.5950 & [5.4236, 5.7698] & 5.5967 & 60 $\pm$ 22 & 75 $\pm$ 28 & 55 $\pm$ 26 & 62 $\pm$ 19 \\
5.9408 & [5.7698, 6.1161] & 5.9429 & 41 $\pm$ 20 & 104 $\pm$ 27 & 14 $\pm$ 32 & 53 $\pm$ 19 \\
6.2860 & [6.1161, 6.4623] & 6.2892 & -11 $\pm$ 21 & 58 $\pm$ 25 & -2 $\pm$ 25 & 12 $\pm$ 18 \\
6.6329 & [6.4623, 6.8085] & 6.6354 & -19 $\pm$ 24 & 76 $\pm$ 28 & 39 $\pm$ 40 & 24 $\pm$ 22 \\
6.9867 & [6.8085, 7.1547] & 6.9816 & 37 $\pm$ 22 & 50 $\pm$ 27 & 1 $\pm$ 34 & 34 $\pm$ 20 \\
7.3250 & [7.1547, 7.5010] & 7.3279 & 62 $\pm$ 22 & 26 $\pm$ 29 & 29 $\pm$ 31 & 45 $\pm$ 20 \\
7.6695 & [7.5010, 7.8472] & 7.6741 & 34 $\pm$ 25 & 65 $\pm$ 29 & 43 $\pm$ 33 & 46 $\pm$ 22 \\
8.0168 & [7.8472, 8.1934] & 8.0203 & 91 $\pm$ 24 & 77 $\pm$ 30 & 39 $\pm$ 31 & 73 $\pm$ 21 \\
8.3635 & [8.1934, 8.5397] & 8.3666 & 82 $\pm$ 24 & 64 $\pm$ 35 & 89 $\pm$ 34 & 80 $\pm$ 23 \\
8.7096 & [8.5397, 8.8859] & 8.7128 & 67 $\pm$ 28 & 16 $\pm$ 34 & -23 $\pm$ 36 & 28 $\pm$ 24 \\
9.0544 & [8.8859, 9.2321] & 9.0590 & 87 $\pm$ 28 & 126 $\pm$ 38 & 22 $\pm$ 35 & 78 $\pm$ 25 \\
9.4019 & [9.2321, 9.5784] & 9.4052 & 42 $\pm$ 28 & 119 $\pm$ 40 & 7 $\pm$ 34 & 49 $\pm$ 25 \\
9.7467 & [9.5784, 9.9246] & 9.7515 & 123 $\pm$ 32 & 136 $\pm$ 38 & 10 $\pm$ 39 & 95 $\pm$ 27 \\
10.0894 & [9.9246, 10.2708] & 10.0977 & 150 $\pm$ 34 & 134 $\pm$ 46 & 42 $\pm$ 45 & 116 $\pm$ 31 \\
10.4320 & [10.2708, 10.6171] & 10.4439 & 33 $\pm$ 41 & 105 $\pm$ 53 & -19 $\pm$ 56 & 40 $\pm$ 37 \\
10.7781 & [10.6171, 10.9633] & 10.7902 & 100 $\pm$ 49 & 251 $\pm$ 61 & 110 $\pm$ 64 & 147 $\pm$ 44 \\
11.1276 & [10.9633, 11.3095] & 11.1364 & 52 $\pm$ 64 & 409 $\pm$ 83 & 27 $\pm$ 96 & 150 $\pm$ 59 \\
11.4733 & [11.3095, 11.6558] & 11.4826 & 268 $\pm$ 72 & -185 $\pm$ 98 & 276 $\pm$ 114 & 143 $\pm$ 69 \\
11.8142 & [11.6558, 12.0020] & 11.8289 & 139 $\pm$ 97 & -17 $\pm$ 116 & 93 $\pm$ 134 & 79 $\pm$ 87 \\
\enddata
\tablecomments{Mean and standard deviation. Inflation ratios are 2.84, 2.04, 1.62, 1.33}
\end{deluxetable*}

\end{document}